\documentclass{aastex63}

\usepackage{amsmath}

\submitjournal{ApJ}

\begin{document}

\title{Thermocline Depth on Water-rich Exoplanets}

\correspondingauthor{Jun Yang}
\email{junyang@pku.edu.cn}
\author[0000-0001-9700-9121]{Yanhong Lai}
\affiliation{Laboratory for Climate and Ocean-Atmosphere Studies, Department of Atmospheric and Oceanic Sciences, School of Physics, Peking University, Beijing 100871, China}
\author[0000-0001-6031-2485]{Jun Yang}
\affiliation{Laboratory for Climate and Ocean-Atmosphere Studies, Department of Atmospheric and Oceanic Sciences, School of Physics, Peking University, Beijing 100871, China}

\begin{abstract}

Water-rich exoplanet is a type of terrestrial planet that is water-rich and its ocean depth can reach tens of to hundreds of kilo-meters with no exposed continents. 
Due to the lack of exposed continents, neither western boundary current nor coastal upwelling exists, and ocean overturning circulation becomes the most important way to return the nutrients deposited in deep ocean back to the thermocline and to the surface ocean.
Here we investigate the depth of the thermocline in both wind-dominated and mixing-dominated systems on water-rich exoplanets using the global ocean model MITgcm. We find that the wind-driven circulation is dominated by overturning cells through Ekman pumping and subduction and by zonal (west--east) circum-longitudinal currents, similar to the Antarctic Circumpolar Current on Earth. The wind-influenced thermocline depth shows little dependence on the ocean depth, and under a large range of parameters, the thermocline is restricted within the upper layers of the ocean. The mixing-influenced thermocline is limited within the upper 10 km of the ocean and can not reach the bottom of the ocean even under extremely strong vertical mixing.  The scaling theories for the thermocline depth on Earth are applicable for the thermocline depth on water-rich exoplanets. However, due to the lack of exposed continents, the zonal and meridional flow speeds are not in the same magnitude as that in the oceans of Earth, which results in the scaling relationships for water-rich exoplanets are a little different from that used on Earth.

\end{abstract}


\section{INTRODUCTION} \label{sec:intro}
Water-rich exoplanet (alternatively called as ocean planet) is a type of proposed terrestrial planet that is covered by a global ocean and without any exposed continent at the surface \citep{kuchner2003volatile,leger2004new,SELSIS2007453,SOTIN2007337,Marcus_2010,2011ApJ...733....2N,Zeng_2014,10.1093/mnras/stw321,Auclair-Desrotour2018,Kite_2018}. The ocean can reach tens of to hundreds of kilometers, which is much deeper than the global-mean depth of the ocean on Earth, $\sim$3 km \citep{stewart2008introduction}. Due to the deep ocean, the pressure at the bottom of the ocean can be extremely high and it results in the occurrence of high-pressure ice at the sea floor. The existence of water-rich exoplanets and their possible bulk compositions have not yet been demonstrated directly by observations, but implied by the mean density derived from observed masses and radii. Their rather larger radii and lower densities compared to other rocky planets (such as Earth) suggest that they may be volatile-rich or water-rich outside the rocky core \citep{Zeng2019}. Water-rich exoplanets may be common in the galaxy, especially around M-type stars \citep{mulders2015stellar,brugger2016possible,Alibert2017}. 

Carbonate-silicate cycle plays an important role in regulating planetary climate and can strongly affect planetary habitability \citep{Walker1981,pierrehumbert_2010}. On water-rich exoplanets, due to the lack of exposed continents and continental weathering, the carbonate-silicate cycle has a weaker dependence on atmospheric CO$_2$, which might make the climate being better at resisting changes in external forcings \citep{Hayworth_2020}. 
\cite{Krissansen_Totton_2021} suggested that the high pressure at the sea floor is unfavorable for magmatic outgassing and seafloor weathering. Considering the temperature dependence of seafloor weathering might improve the habitability of water-rich exoplanets \citep{Abbot2012}. 
Moreover, the carbon partition between the atmosphere and ocean might play a negative role in stabilizing the climate on water-rich exoplanets, due to the decreased capability of the ocean to absorb carbon dioxide with temperature increasing \citep{kitzmann2015unstable}.

Ocean circulation also can affect the habitability of water-rich exoplanets. Different from other terrestrial exoplanets with shallow oceans like Earth, there is no subtropical or subpolar gyres, boundary currents, and coastal upwelling motions on water-rich exoplanets due to the lack of exposed continents. In particular, coastal upwelling plays a significant role in the upward transport of nutrients from the deep ocean to the upper layers on Earth, which can directly affect the biological activity in the ocean \citep{Hutchings1995}. \cite{Olson_2020} demonstrated the importance of considering the wind-driven upwelling for the nutrient transport and biological activity on exoplanets with exposed continents and land-sea contrasts. However, there is no coastal upwelling on water-rich exoplanets. 
The meridional overturning circulation can also transport heat poleward, regulate the mean climate and thereby play a significant role in affecting the planetary habitability. 
\cite{Vallis2009} suggested that the effect of different ocean depths on the overturning circulation and the meridional ocean heat transport is limited, as long as the ocean depth is deeper than the main thermocline (a layer where the vertical gradient of temperature is large and is not subject to seasonal variability; \citealt{stewart2008introduction}). However, the maximum ocean depth they tested is 4 km, simulations with deeper oceans have not been investigated. 

The thermocline can strongly influence buoyancy, circulation, and the exchange of oxygen, carbon, heat and other nutrients between the upper layer and the lower layer \citep{ zelle2004relationship,cantin2011effects,giling2017thermocline}. The thermocline is a layer of water where the temperature decreases with depth more rapidly and the vertical temperature gradient is larger than it does in the layers above or below \citep{pedlosky2006history,stewart2008introduction,fiedler2010comparison}. The thermocline on Earth can be divided into two kinds, one is main thermocline, the other is seasonal thermocline that varies with the seasons \citep{stewart2008introduction}.
The thermocline separates the warm mixed layer (turbulent and properties uniformly distributed vertically) from the cold deep water, and the mixing between the upper and lower layers is inhibited by the large gradient. Due to the lack of inter-region mixing, oxygen content below the thermocline rapidly depletes with depth increasing, since organisms utilize it and there is no source below the thermocline where there is no sunlight \citep{thistle2003deep}. Given that density is partly determined by temperature, and the vertical density gradient within the thermocline is also larger than the upper mixed layer and the lower layer. 

Here we firstly explore the depth of the thermocline under different ocean depths. Then, the sensitivity of the thermocline depth in wind-dominated system and mixing-dominated system to various planetary and oceanic parameters is investigated.  
Section \ref{sec:model} presents our model and experimental designs. Section \ref{sec:results} presents our results, including the wind-driven circulation (Section \ref{subsec:wind-driven}) and the thermal-driven circulation (Section \ref{subsec:thermal-driven}). We summarize and discuss the results in Section \ref{sec:summary}. 

\section{MODEL DESCRIPTIONs AND EXPERIMENTAL DESIGNs} \label{sec:model}
The global ocean model used in this study is the Massachusetts Institute of Technology general circulation model (MITgcm; \citealt{https://doi.org/10.1029/96JC02775,https://doi.org/10.1029/96JC02776}). By default, we simulate one planet with a global ocean that has a uniform depth of 40 km and has no any continental barrier in the zonal direction. In meridional direction, the ocean covers from 80$^{\circ}$S to 80$^{\circ}$N; higher latitudes are not simulated and solid walls are placed at the southern and northern boundaries of our domain in order to avoid grid cell convergence at the polar points. The primitive equations in spherical coordinates we used are written as: 
\begin{equation}
    {{D u} \over {D t}} - {u v \over a} tan\varphi - f v + {1 \over {\rho_c a cos \varphi}} {\partial p' \over \partial \lambda} - {{\nabla}_h \cdot (A_h {\nabla}_h u )} - {\partial \over \partial z} (A_z {\partial u \over \partial z}) = F_u,
\label{equat1}
\end{equation}
\begin{equation}
    {{D v} \over {D t}} + {u^2 \over a} tan\varphi + f u + {1 \over \rho_c a} {\partial p' \over \partial \varphi} - {{\nabla}_h \cdot (A_h {\nabla}_h v )} - {\partial \over \partial z} (A_z {\partial v \over \partial z}) = F_v,
\label{equat2}    
\end{equation}
\begin{equation}
     {\partial p' \over \partial z} = -\rho' g,
\label{equat3}    
\end{equation}
\begin{equation}
    {{\nabla}_h \cdot \vec{u} + {\partial w \over \partial z}} = 0,
\label{equat4}    
\end{equation}
\begin{equation}
    \rho' = \rho (\theta,S,p_0(z))-\rho_c,
\label{equat5}
\end{equation} 
\begin{equation}
    {{D \theta} \over {D t}}  - {{\nabla}_h \cdot (k_h {\nabla}_h \theta )} - {\partial \over \partial z} (k_v {\partial \theta \over \partial z})= F_{\theta}, 
\label{equat6}    
\end{equation}
\begin{equation}
    {{D S} \over {D t}} - {{\nabla}_h \cdot (k_{hS} {\nabla}_h S )} - {\partial \over \partial z} (k_{vS} {\partial S \over \partial z}) = F_S,
\label{equat7}
\end{equation}
where $\vec{u}$ is the horizontal flow vector, and $u$ and $v$ are its zonal and meridional components, respectively; $w$ is the vertical current speed;  $\theta$ and $S$ are the potential temperature and salinity, respectively; the pressure field is splitted into a reference function of height $p_0(z)$ and a perturbation term $p'$; $\rho$ is the seawater density, $\rho_c$ is the reference density (1035 kg\,m$^{-3}$ in our simulations), and $\rho'$ is the density variation relative to the reference value;  $g$ is the gravitational acceleration; $\lambda$ is the longitude; $\varphi$ is the latitude; $z$ is the vertical distance from the surface and is negative for seawater below the surface; $a=R+z$ is the distance from the center of the Earth, where $R$ is the planetary radius; generally $z \ll R$ and the approximation $a \approx R$ is used in models; ${D \over Dt} = {\partial \over {\partial t}}+ \vec{u} \cdot {\nabla}_h + w \, {\partial \over \partial z}$ is the total derivative, where ${\nabla}_h = {1 \over {acos \varphi}} {\partial \, \over \partial \lambda} \, \vec{i} +{1 \over a} {\partial \, \over {\partial \varphi}} \, \vec{j}$ ( $\vec{i}$ and $\vec{j}$ are the zonal and meridional unit vectors, respectively); 
$A_h$ and $A_z$ are horizontal and vertical viscosities; $k_h$ and $k_v$ are horizontal and vertical diffusivities for potential temperature; $k_{hS}$ and $k_{vS}$ are horizontal and vertical diffusivities for salinity; the Coriolis parameter ($f$) is defined as $2 \Omega sin \varphi$, where $\Omega$ represents the planetary rotation rate; 
$F_u$, $F_v$, $F_{\theta}$, and $F_S$ are zonal wind stress, meridional wind stress, sea surface temperature forcing, and surface salinity forcing, respectively.

\begin{figure*}[ht]     
\centering
\includegraphics[scale=0.45]{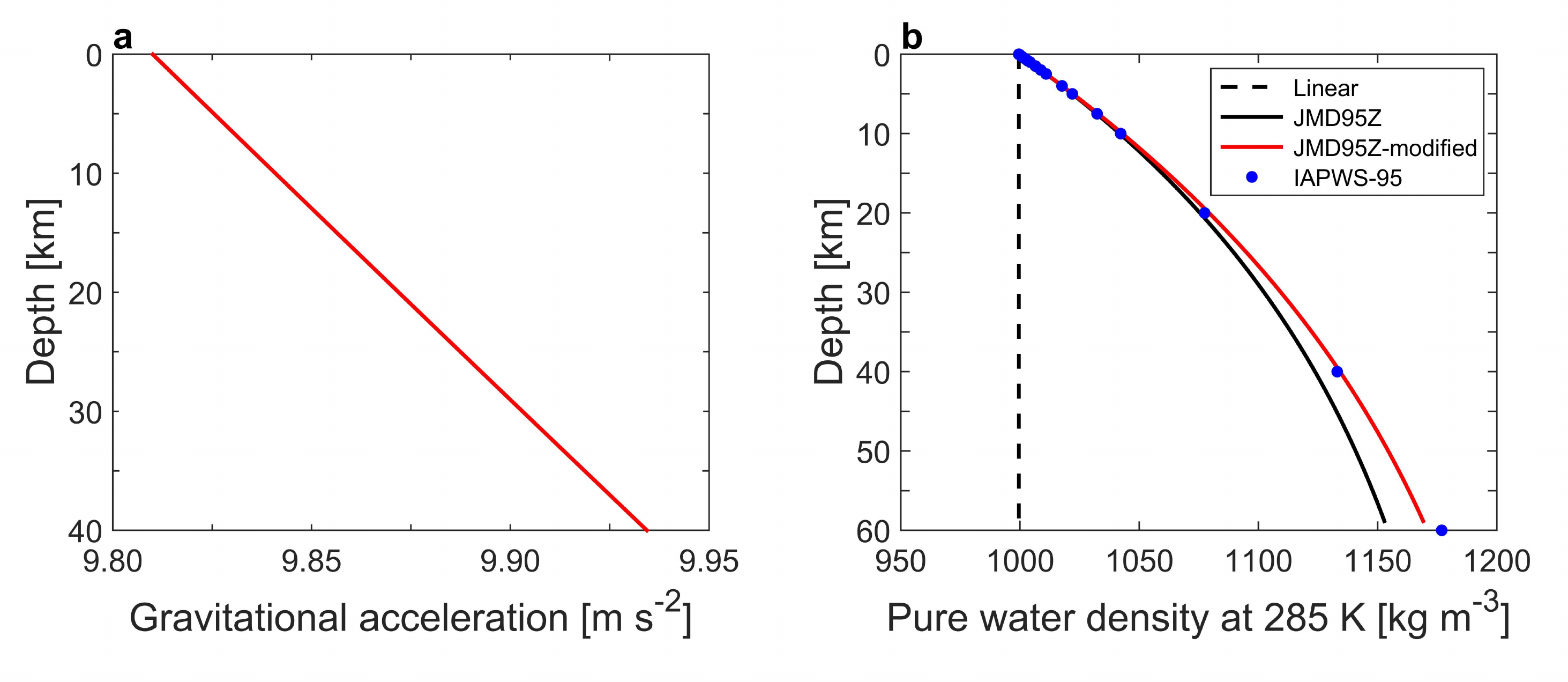}
\caption{(a) Gravitational acceleration as a function of depth under Earth-like mass and radius but with an ocean depth of 40 km. 
(b) Comparisons of the depth dependence of density for pure water between four different equations of state (EoS),
including a linear EoS (density varies with potential temperature and salinity linearly but does not depend on pressure), the origional `JMD95Z', modified 'JMD95Z' under a conserved potential temperature of 285 K, and `IAPWS-95' under a fixed in-situ temperature of 285 K. Below we show that different EoSs will not significantly affect the simulation results.\label{fig1}}
\end{figure*}

For simulations with shallow oceans, $g$ is set to be constant throughout the entire ocean. While for the simulations with a 40-km-deep ocean, the depth dependence of $g$ may be needed to be considered (Figure \ref{fig1}(a)), given by:
\begin{equation}
    g(z) = {GM \over {(R+z)}^2 }= {g_0 {({R \over {R+z}})}^2},
    \label{equat8}
\end{equation}
where $G$ is the gravitational constant, $M$ is the planetary mass, and $g_0$ is the gravitational acceleration at the sea surface and is set to be 9.81 m\,s$^{-2}$. It should be noted that gravity increases with depth when the core is significantly denser than water such as for terrestrial planets, and gravity should decrease with depth if the planet is homogeneous in density \citep{dragoni2020gravity}.

\begin{table}
\caption{Planetary and Oceanic Parameters in the Control Experiment \label{tab1}}
\centering
\begin{tabular}{l l}
\hline
Parameter  & Value  \\
\hline
Radius (R) & 6371 km\\
Rotation rate ($\Omega$) & 7.27 $\times 10^{-5}$ s$^{-1}$\\
Surface gravity (g$_0$) & 9.81 m\,s$^{-2}$\\
Ocean depth & 40 km\\
Salinity (S) & 34.7 g\,kg$^{-1}$\\
Horizontal viscosity (A$_h$) & 1.2 $\times 10^{6}$ m$^2$\,s$^{-1}$\\
Vertical viscosity (A$_z$) & 10$^{-3}$ m$^2$\,s$^{-1}$\\
Vertical diffusivity (k$_v$) & $10^{-4}$ (surface) and $10^{-5}$ m$^2$\,s$^{-1}$ (interior)\\
\hline
\multicolumn{2}{l}{}
\end{tabular}
\end{table}

In the control experiment, most of the planetary and oceanic parameters resemble present-day Earth (Table \ref{tab1}). By default, $R$ is 6371 km, $\Omega$ is 7.27 $\times$ 10$^{-5}$ s$^{-1}$, $g_0$ is 9.81 m\,s$^{-2}$, and the ocean depth is 40 km. $A_h$ and $A_z$ are 1.2 $\times$ 10$^6$ and 10$^{-3}$ m$^2$\,s$^{-1}$, respectively. $k_v$ is 10$^{-5}$ m$^2$\,s$^{-1}$ in the interior of the ocean and 10$^{-4}$ m$^2$\,s$^{-1}$ at the sea surface due to the effect of wind stress mixing. $k_h$ is zero and the mixing of tracer properties along isopycnal is represented by the Gent-McWilliams scheme (GM, \citealt{IsopycnalMixinginOceanCirculationModels}) with the Redi eddy parameterization \citep{OceanicIsopycnalMixingbyCoordinateRotation}. In this study, the surface salinity forcing is not considered. Thus, for simplicity $S$ is set to be 34.7 g\,kg$^{-1}$ everywhere and will not evolve with time, and both $k_{hS}$ and $k_{vS}$ are not used. 

The equation of state (EoS) reflects the dependence of density on potential temperature, salinity, and pressure. In MITgcm, the nonlinear EoS `JMD95Z' \citep{MinimalAdjustmentofHydrographicProfilestoAchieveStaticStability} is relatively accurate when the ocean depth is shallower than $\sim$10 km. Below 10 km, the accuracy of `JMD95Z' is less. To calculate the density in a 40-km-deep ocean, `JMD95Z' is modified following the `IAPWS-95' formulation \citep{WagnerandPrub1995} that covers a validity range for temperatures from 0 $^{\circ}$C to 1000 $^{\circ}$C and pressures up to 10$^{9}$ Pa. After modification, the accuracy of density is better guaranteed, as shown in Figure \ref{fig1}(b). The detailed relationship between the density of seawater and salinity, potential temperature, and pressure under the nonlinear EoS, `JMD95Z', and modified `JMD95Z', is given in Appendix \ref{sec: nonlinear_eos}. 

\begin{figure*}[ht]     
\centering
\includegraphics[scale=0.55]{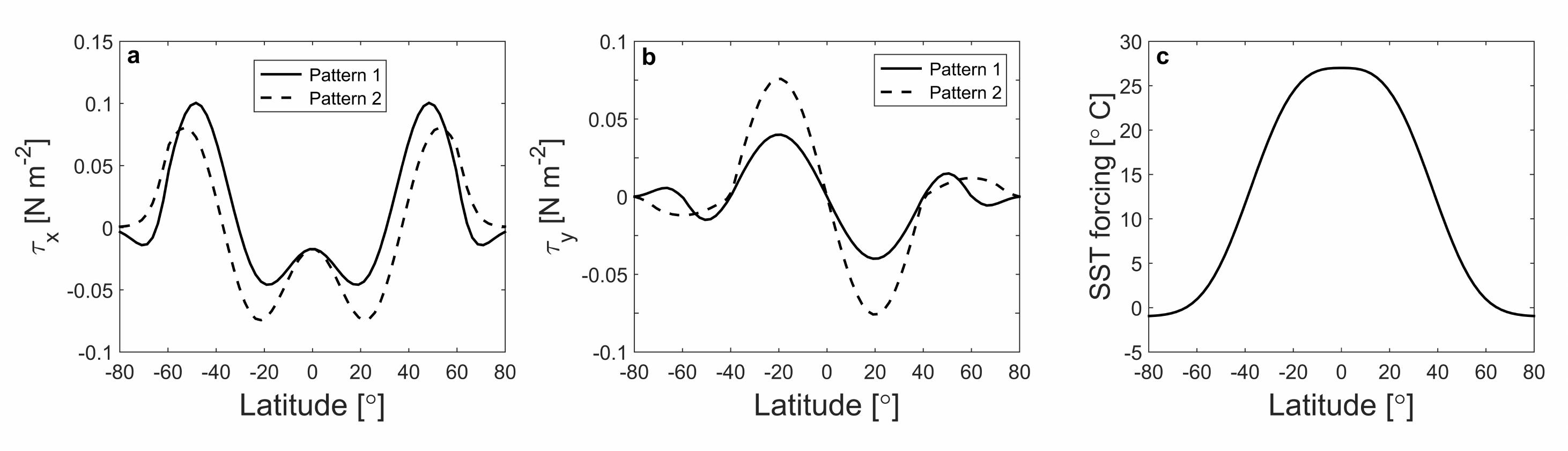}
\caption{External forcings imposed at the sea surface. (a) Zonally-symmetric zonal wind stress in N\,m$^{-2}$ (positive: eastward; negative: westward); (b) zonally-symmetric meridional wind stress in N\,m$^{-2}$ (positive: northward; negative: southward). Solid line: the spatial pattern of wind stress that resembles current Earth; dashed line: a different spatial pattern, which will be tested in Section \ref{sec:results}. (c) Zonally-symmetric sea surface temperature (SST) forcing in $^{\circ}$C as a restoring boundary condition.\label{fig2}}
\end{figure*}

The ocean circulation is forced by the atmosphere through the wind stresses and sea surface temperatures, but the full interaction between the atmosphere and ocean is not considered in this study. Zonal and meridional wind forcings, $F_u$ and $F_v$, are given by $\tau_x \over {\rho_c \Delta z_s }$ and ${\tau_y \over {\rho_c \Delta z_s }}$, respectively, where $\tau_x$ and $\tau_y$ are zonally-symmetric zonal and meridional wind stresses, and $\Delta z_s$ is the depth of the surface layer of the ocean. By default, the wind stress resembling the present-day Earth is imposed (solid lines in Figure \ref{fig2}(a) \& (b)). The sea surface temperatures are set to be restored to a zonally-symmetric profile ($\theta_{\ast}$) similar to current Earth (Figure \ref{fig2}(c)) with a relaxation timescale ($\tau_{\theta}$) being equal to 30 Earth days, given as $F_{\theta} = -{{1 \over \tau_{\theta}} (\theta - \theta^{\ast})}$.
Both the wind stress and sea surface temperature forcings are applied only in the surface layer of the model. 

We perform series of sensitivity simulations to investigate the depth of the  thermocline under different planetary and oceanic parameters, including ocean depth, EoS for seawater, depth dependence of gravitational acceleration, parameterization scheme for mesoscale eddies, viscosity, wind stress, vertical diffusivity, and planetary rotation rate. We briefly outline our procedures for these simulations below and Table \ref{tab2} summarizes the corresponding setup and the range for each parameter. 

\begin{table}
\caption{Summary of the Main Simulations Performed in This Study\label{tab2}}
\centering
\begin{tabular}{lp{1cm}p{11cm}p{5cm}}
\hline
Group  & Runs & Experimental design \\
\hline
Control (2D) & 1 & The planet is covered with a global ocean with a uniform depth of 40 km. Modified `JMD95Z' equation of state is used. The depth dependence of gravitational acceleration following Equation (\ref{equat7}) and Figure \ref{fig1}(a) is considered. The horizontal diffusivity is zero and the GM scheme is used. Planetary and oceanic parameters are Earth-like except the ocean depth (Table \ref{tab1}). \\
Ocean depth & 2 & Same as ``Control'' except the ocean depth is 5 or 10 km.\\
Wind stress & 4 &  Same as ``Control'' except the wind stresses are 0.5, 2, or 4 times that of the control case, or its spatial pattern is varied (Figure \ref{fig2}). \\
Rotation rate & 20 &  Same as ``Control'' except the planetary rotation rate is 1/30, 1/15, 1/13, 1/11, 1/10, 1/7, 1/5, 1/3, 1/2, 2, 4, 6, 8, 12, or 24 times that in the control case.\\
Vertical diffusivity & 14 &  Same as ``Control'' except the interior vertical diffusivity is 10$^{-6}$, 2\,$\times$\,10$^{-6}$, 4\,$\times$\,10$^{-6}$, 2\,$\times$\,10$^{-5}$, 4\,$\times$\,10$^{-5}$, 6\,$\times$\,10$^{-5}$, 8\,$\times$\,10$^{-5}$, 10$^{-4}$, 2\,$\times$\,10$^{-4}$, 4\,$\times$\,10$^{-4}$, 6\,$\times$\,10$^{-4}$, 8\,$\times$\,10$^{-4}$, 10$^{-3}$, or 2\,$\times$\,10$^{-3}$ m$^2$\,s$^{-1}$. And, wind stress forcing is not included.\\
Equation of state (EoS) & 2 &  Same as ``Control'' except the EoS used is linear with a thermal expansion coefficient of 2\,$\times$\,10$^{-4}$ m$^2$\,s$^{-1}$ or the nonlinear original `JMD95Z' (see Figure \ref{fig1}(b)).\\
Gravity & 1 & Same as ``Control'' except the gravity acceleration is set to be 9.81 m\,s$^{-2}$ throughout the entire ocean and its depth dependence is not considered. \\
GM scheme & 1 &  Same as ``Control'' except the GM scheme is turned off.\\
Viscosity & 2 &  Same as ``Control'' except the horizontal and vertical viscosities are decreased by a factor of 2 or 10 at the same time. \\
Three-dimensional (3D) & 7 &  Same as ``Control'' except the zonal direction is also included. \\
\hline
\multicolumn{2}{l}{}
\end{tabular}
\end{table}

By default, the ocean depth is set to be 40 km. In order to compare that with shallow oceans, we do two experiments with ocean depths of 5 km and 10 km. In the shallow ocean simulations, the variation of gravitational acceleration with depth is not considered. 

Wind forcing and vertical mixing are two main factors that make the ocean circulation driven by surface density/heat forcing not be restricted within a thin layer at the surface but reach deeper \citep{vallis2019essentials}. The magnitude of wind stresses ranging from 0.5 to 4 times that in the control experiment is investigated. Generally, the magnitude of wind stresses is determined by atmospheric density, surface wind speed, and drag coefficient \citep{stewart2008introduction}. \cite{Olson_2020} showed that due to the opposite influences of surface pressure on atmospheric density and on surface wind speed, the wind stresses only increase by a factor of two when the surface pressure is increased by a factor of ten.
Also, a case with stronger and wider easterly winds and equatorward winds in the tropics is carried out (see the dashed lines in Figure \ref{fig2}(a) \& (b)), which considers wider Hadley cells induced possibly by a slower planetary rotation rate.  

The strength of mixing is quantified as vertical eddy diffusivity, which can range from about 10$^{-5}$ (background level in the ocean interior) to 10$^{-3}$ m$^2$\,s$^{-1}$ (enhanced mixing due to such as bottom topography) on Earth \citep{GlobalPatternsofDiapycnalMixingfromMeasurementsoftheTurbulentDissipationRate}. \cite{si2021planetary} use simple scalings and obtain the possible strength of vertical mixing on asynchronous rotating habitable exoplanets around M-dwarfs, which is roughly 100 times the mean value on Earth. Interior diffusivities varying between 10$^{-6}$ and 2\,$\times$\,10$^{-3}$ m$^2$\,s$^{-1}$ are tested in this study. When the mixing is strong, the equilibrated system is no longer wind-dominated but mixing-dominated.

Planetary rotation rates of exoplanets may range from several hours to several hundred Earth days \citep{Akeson_2013,Impey2013}. The rotation rate can directly affect the strength of Ekman pumping and subduction and then affect the wind-driven circulation \citep{vallis2019essentials}. 
Planetary rotation rates ranging from 1/30 to 24 times that of Earth under both low-diffusivity (10$^{-5}$ m$^2$\,s$^{-1}$) and high-diffusivity (10$^{-3}$ m$^2$\,s$^{-1}$) cases are examined in this study.

The GM scheme is always used to parameterize the effect of meso-scale eddies on tracer transports when using a non eddy-resolving ocean circulation model.  Studies on Earth show that the GM scheme can effectively resolve the effect of meso-scale eddies and exert an evident impact on the temperature and the thermocline \citep{danabasoglu1995sensitivity}. In mid-latitudes, where the meridional temperature gradients are large and baroclinic eddy activities are strong, the usage of GM scheme instead of a specified horizontal diffusion can greatly reduce the temperature bias between models and observations. 
One simulation with GM scheme turned off is carried out.

To verify the influence of the modified EoS on the thermocline depth, we also run two simulations using a linear EoS with a thermal expansion coefficient of 2$\times$10$^{-4}$ m$^2$\,s$^{-1}$ or using the original ‘JMD95Z’ (Figure \ref{fig1}(b)). The effect of excluding the depth dependence of gravitational acceleration in 40-km-deep ocean is also tested. The horizontal and vertical viscosities decreased by a factor of 2 and 10 at the same time are also tested due to their possible effects on the thermocline depth.

\begin{figure*}[ht]     
\centering
\includegraphics[scale=0.55]{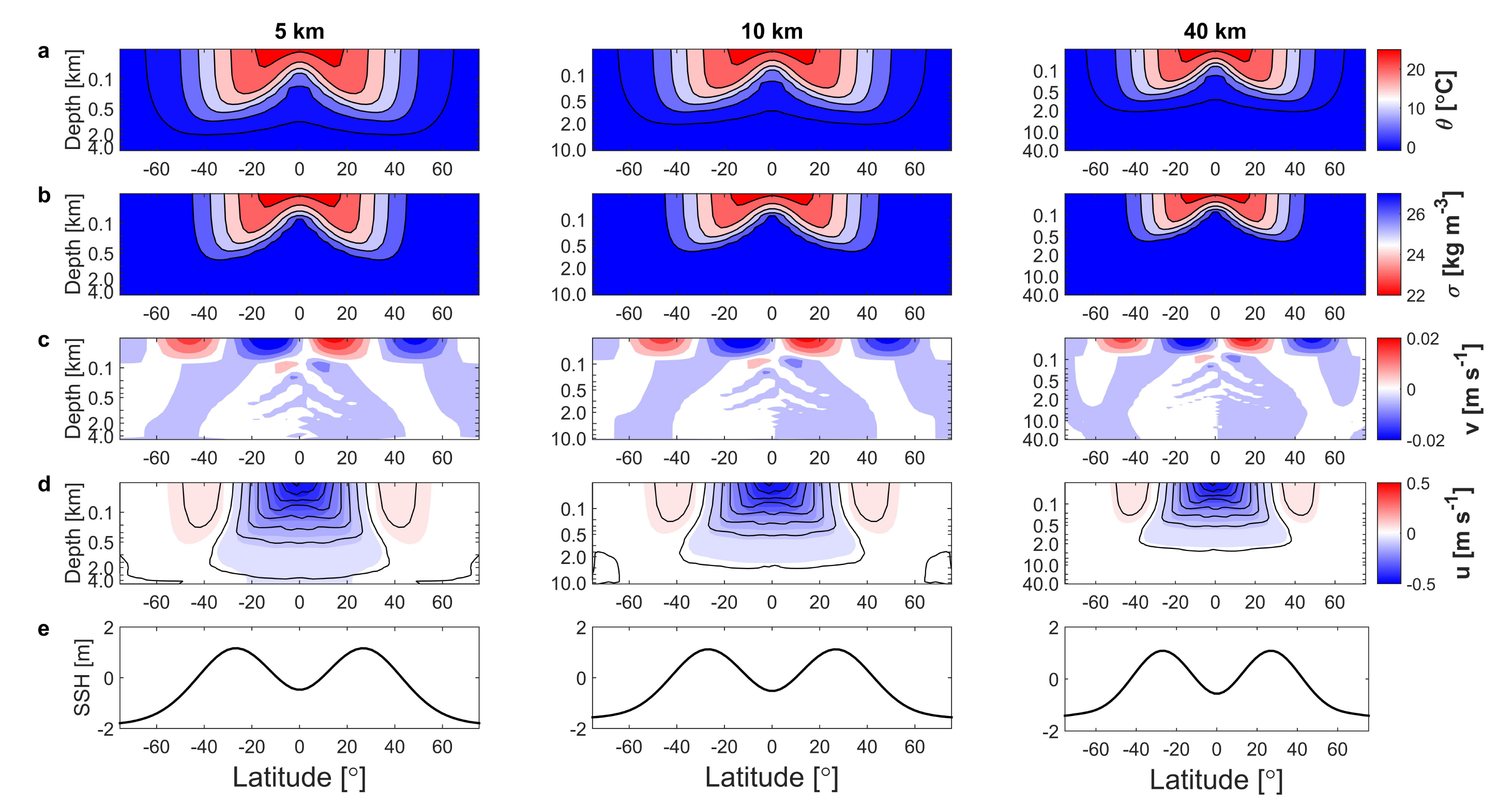}
\caption{Equilibrated fields of the three 2D simulations with different ocean depths. Left: the ocean depth is 5 km; middle: 10 km; right: 40 km. (a) Potential temperature ($^{\circ}$C); (b) potential density (kg\,m$^{-3}$, a value of 1000 kg\,m$^{-3}$ has been subtracted); (c) meridional current (m\,s$^{-1}$); (d) zonal current (color shading) and the corresponding thermal-wind balance current (contour lines) (m\,s$^{-1}$); (e) sea surface height (SSH, m). Note the y-axis in (a)-(d) is nonlinear. \label{fig3}}
\end{figure*}

Due to the zonal symmetry of the large-scale circulation on water-rich exoplanets, simulation results using zonally-symmetric two-dimensional (2D, y--z) models are almost the same to that of three-dimensional (3D) models (see Section \ref{sec:results} below). Thus, 2D models are used in most of our simulations due to computational resource limitations.

We do not consider any topography at the bottom of the ocean, and a flat-bottomed ocean is used. But, a linear bottom drag is included at the sea floor with a linear coefficient of 10$^{-3}$ m$^2$\,s$^{-1}$. Horizontal resolution of the model is 2.25$^{\circ}\times2.25^{\circ}$ in longitude and latitude, respectively. In the vertical, we use 70 unequally spaced levels for the 40-km-deep ocean. The vertical resolution increases nonlinearly from 20 m at the sea surface to 665 m at the sea bottom.
When simulating the 5-km-deep ocean and 10-km-deep ocean, we employ 15 and 24 layers in the vertical, respectively. Our simulations are initialized from a state of rest and are integrated until a statistical equilibrium is reached. For most of our cases, the system achieves quasi-equilibrium within about 15,000 Earth years. If not mentioned, the results shown below are mean states obtained by averaging over the last 1000 years of each integration.

\section{RESULTS} \label{sec:results}

\subsection{Thermocline Depth in Wind-dominated System \label{subsec:wind-driven}}

\begin{figure*}[ht]     
\centering
\includegraphics[scale=0.5]{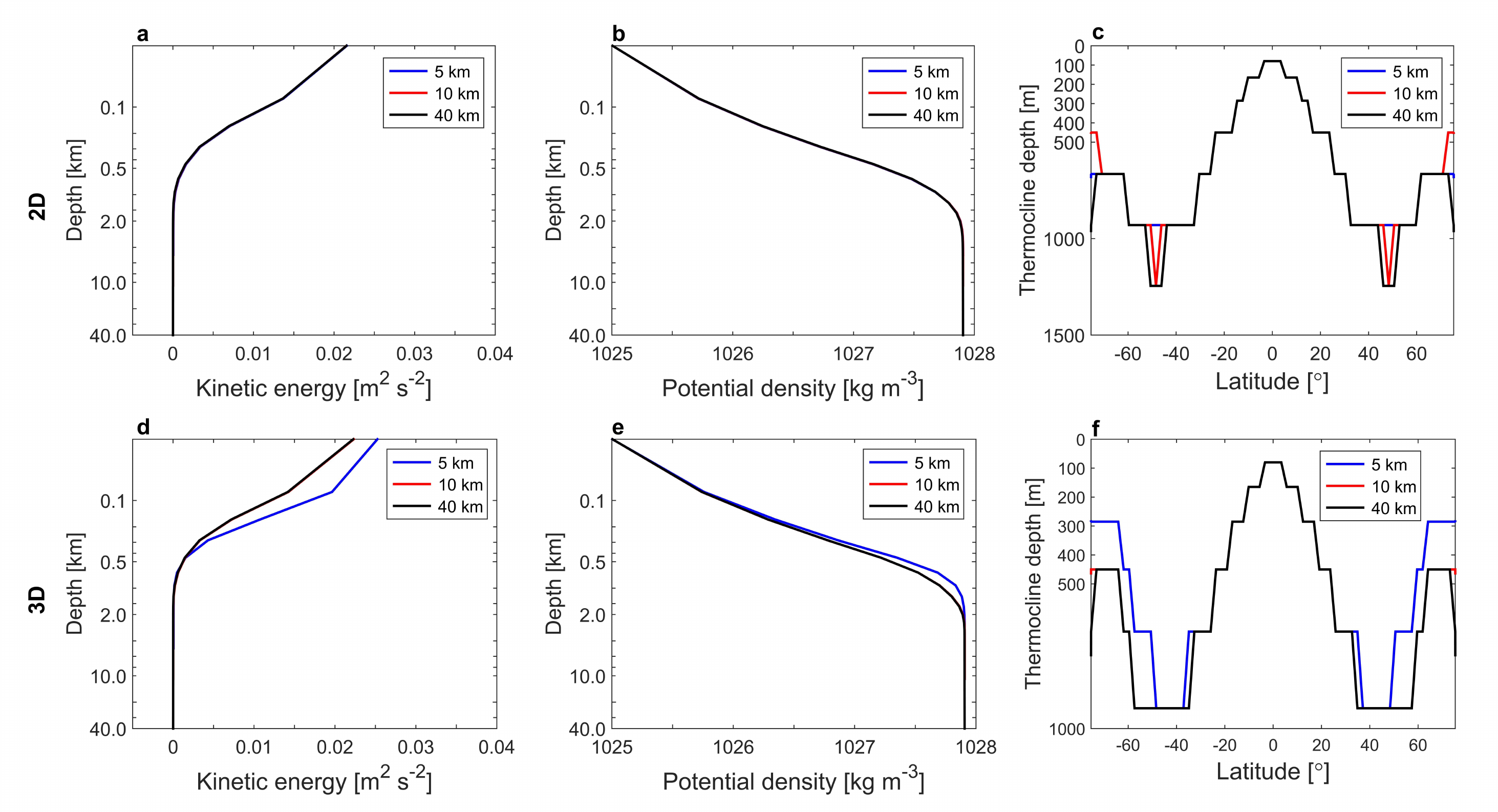}
\caption{Sensitivity of the thermocline depth to ocean depth in 2D models (top row) and in 3D models (bottom row). (a) \& (d) Global-mean kinetic energy profiles (m$^2$\,s$^{-2}$); (b) \& (e) global-mean potential density profiles (kg\,m$^{-3}$); (c) \& (f) thermocline depth (m) as a function of latitude. In (a) and (b), three lines are completely overlapped with each other. In (d), (e), and (f), the red line is overlapped by the black line. In (a), (b), (d), and (e), the y-axis is nonlinear. The results of 2D and 3D simulations are quite similar.\label{fig4}}
\end{figure*}

We find that the thermocline depth on water-rich exoplanets with deep oceans is as shallow as that on planets with shallow oceans (Figure \ref{fig3} $\&$ \ref{fig4}). In the three cases with different ocean depths (5, 10, and 40 km), cold water upwells near the equator and $\pm$\,60$^{\circ}$ latitudes, and warm water at the surface sinks near $\pm$\,30$^{\circ}$ latitudes, which results in lower temperatures in upwelling regions and higher temperatures in downwelling regions in the upper ocean (Figure \ref{fig3}(a)). Thus, the temperature field exhibits a bowl-shaped pattern. The potential density field is similar to the potential temperature field (Figure \ref{fig3}(b)) due to the spatially uniform salinity used in this study. In the meantime, there are poleward flows in the tropics and equatorward flows in the mid-latitudes in the surface layer (Figure \ref{fig3}(c)). Thus, the wind-driven meridional overturning circulation is clockwise in the tropics and anti-clockwise in the mid-latitudes. Ekman pumping and subduction induced by the imposed wind stresses can account for the equilibrated wind-driven overturning circulation \citep{vallis2019essentials}, and the vertical velocity at the bottom of the Ekman layer can be given as
\begin{equation}
W_E = {1\over {\rho_c}} {curl_z{\tau\over f}},
\label{equat9}
\end{equation}
where $\tau$ is the imposed wind stress and $curl_z$ takes its curl in the vertical. Equation (\ref{equat9}) shows that wind-driven Ekman pumping and subduction induce upwelling of cold water where the wind stress curl is positive in the northern hemisphere (negative in the southern hemisphere) and downwelling of warm water where the wind stress curl is negative in the northern hemisphere (positive in the southern hemisphere). The simulation results (Figures \ref{fig3} \& \ref{fig5}) are consistent with Equation (\ref{equat9}). 

The meridional currents at the sea surface are in the balance between the Coriolis force and the zonal wind stresses Appendix \ref{sec: momentum_budget}), so the merifional flow must be poleward in which the zonal wind stress is easterly and equatorward in which the zonal wind stress is westerly in the northern hemisphere (Figure \ref{fig3}(c)). 

Equation (\ref{equat9}) also suggests that there is no relationship between the ocean depth and the strength of Ekman pumping (subduction), i.e., the strength of wind-driven overturning circulation may be the same under different ocean depths, which can be found in Figure \ref{fig3}. The thermocline depth also exhibits little dependence on the ocean depth (Figure \ref{fig4}(a)-(c)). In all cases, the thermocline depth is restricted within the upper 2 km. Below 2 km, both the potential temperature and potential density are nearly uniform.
Quantitative calculations based on the e-folding depth of potential density \citep{bryan1987parameter} shows that the thermocline depths in the three cases with different ocean depths are nearly the same (Figure \ref{fig4}(c)). In meridional direction, the thermocline is shallow near the equator where the cold water is brought up and the vertical temperature (or potential density) gradient is relatively large, and the thermocline is deep at the mid-latitudes where the warm water sinks and the vertical temperature gradient is relatively small.

There is also no obvious difference in the sea surface height (SSH, Figure \ref{fig3}(e)) and zonal current fields (Figure \ref{fig3}(d)) among the three cases with different ocean depths. The latitudinal distribution of SSH is dominated by the effect of seawater temperature, i.e., SSH is high where the temperature of the underlying ocean is high and is low where the temperature of the underlying ocean is low. As a result, there is a SSH minimum near the equator due to the upwelling of cold water and a SSH maximum near 30$^\circ$ because of the downwelling of warm water. 
The meridional gradient of SSH determines the meridional gradient of seawater pressure and thereby the speed of zonal currents through geostrophic balance both at the ocean surface and in the ocean interior (Figure \ref{figB1} in Appendix \ref{sec: momentum_budget}). Thus, the ocean flows are westward in the tropics and eastward in the mid-latitudes (Figure \ref{fig3}(d), colors). With depth increasing, the speed of the zonal currents decreases, following thermal-wind balance (contour lines in Figure \ref{fig3}(d); \citealt{vallis2019essentials}).

As the ocean becomes deeper, the zonal flows become closer to the thermal-wind balance, as the calculated thermal-wind currents do not coincide with the zonal flows in the 5-km-deep ocean case as good as that in the 40-km-deep ocean case (Figure \ref{fig3}(d)). This is probably because the influence of the bottom drag on the zonal flows becomes weaker as the ocean is deeper. 

The thermocline depth  with varying ocean depths is also tested using 3D configuration. Zonally-symmetric external forcings as the solid lines in Figure \ref{fig2} are employed in the 3D simulations. Due to the axisymmetric external forcings and to the lack of exposed continents, the results of the 3D simulations are quite similar to the 2D simulation results (see Figures \ref{fig4}). It is worth noting that there is baroclinic instability within approximately 30$^{\circ}$S--30$^{\circ}$N in the 3D simulations, while there is not in the 2D simulations. Despite the existence of baroclinic eddy activities, its effect on potential temperature is quite limited. For details, see Appendix \ref{sec: baroclinic_insta}.
Given that the consistency between the results of 3D and 2D simulations, we hereafter choose to use 2D configuration to investigate the relationship between the depth of ocean circulation and other parameters in order to save computational sources.

\subsubsection{The effect of varying wind stresses\label{subsubsec:wind}}

\begin{figure*}[ht]     
\centering
\includegraphics[scale=0.53]{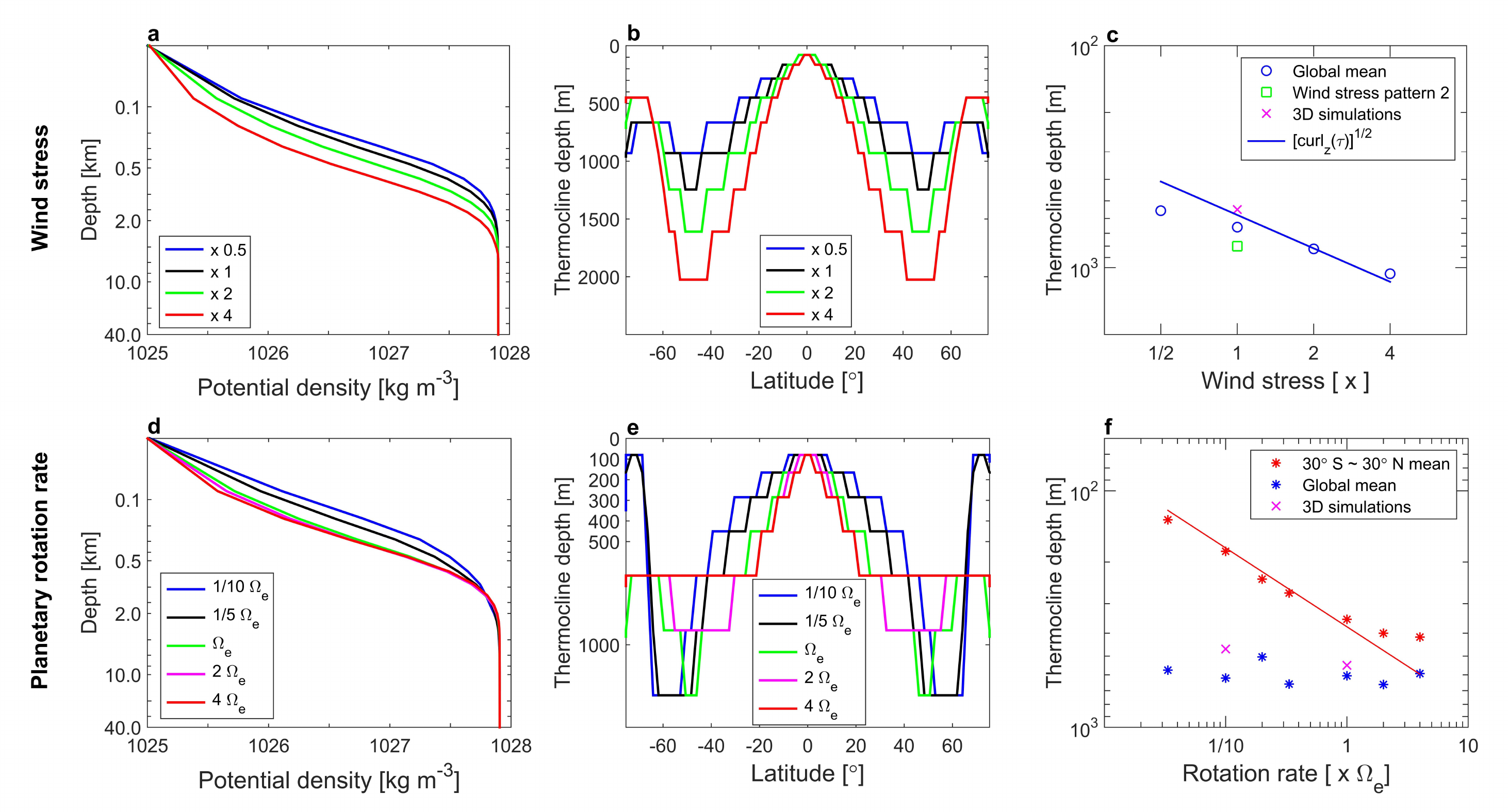}
\caption{Sensitivity of the thermocline depth to wind stress (upper panels) and planetary rotation rate (lower panels). Upper panels: (a) global-mean potential density profiles; (b) thermocline depth as a function of latitude; (c) comparisons between numerical results and the theoretical scaling relation (Equation (\ref{equat17})): blue points are global-mean thermocline depth of numerical results, blue line shows the scaling relation, purple crosses are results of 3D simulations, and green square is the global-mean thermocline depth of a simulation in which the spatial pattern of wind stress is as the dashed lines in Figure \ref{fig2}. Note the y-axis in (a) and (c) is nonlinear. Lower panels: same as the upper panels, but for different planetary rotation rates, and the red points are regional-mean thermocline depth of numerical results, the red line shows the scaling relation Equation (\ref{equat17}), the purple crosses are regional-mean and global-mean thermocline depths of 3D numerical results. \label{fig5}}
\end{figure*}

The wind-driven overturning circulation becomes stronger and the wind-influenced thermocline reaches deeper as the magnitude of the wind stress curl increases (Figure \ref{fig5}(a)-(c)). As Equation (\ref{equat9}) suggests, larger wind stress curl could induce stronger Ekman pumping and then result in stronger overturning circulation. Global-mean potential density profiles under different strengths of wind stress indicate that stronger wind forcings correspond to a deeper thermocline (Figure \ref{fig5}(a)), despite that the changes of the thermocline depth are highly latitude-dependent (Figure \ref{fig5}(b)). With stronger wind forcings, the upwelling motion becomes stronger and then brings colder water to the upper ocean, which results in larger vertical temperature gradients and a shallower thermocline; meanwhile, the downwelling motion in mid-latitudes is stronger and pumps more warm water deeper to the interior ocean, which results in weaker vertical temperature gradients and a deeper thermocline (Figure \ref{fig5}(b)). Overall, the global-mean depth of the thermocline increases with wind forcing, which extends from about 0.6 km to 1.0 km when the wind forcing is increased by a factor of 4 (Figure \ref{fig5}(c)). The influence of varying the spatial pattern of imposed wind stress on the thermocline depth is limited (green square in Figure \ref{fig5}(c)), possibly due to the limited changes of the wind stress curl (Figure \ref{fig2}(a) $\&$ (b)).

A simple scaling for the depth of the wind-influenced thermocline can be derived from thermal wind balance and zonally-symmetric mass continuity under the Boussinesq approximation \citep{Welander1971b,bryan1987parameter,vallis2000large},
\begin{equation}
    f{\partial u \over \partial z}=-{\partial b \over \partial y},
\label{equat10}
\end{equation}
\begin{equation}
    {\partial v \over \partial y}+{\partial w \over \partial z}=0,
\label{equat11}
\end{equation}
where $b=-{\rho' \over \rho_c} g$ is the buoyancy. $b=b'(y,z,t)+\overline{b(z)}$ is composed of two parts: $\overline{b}$ is a reference field and only depends on depth; $b'$ is the perturbation term that varies both in time and space. 
If considering a linear EoS and excluding the salinity effect, we have $\rho =\rho_c (1-\alpha \delta T)$ and $b'=\alpha \delta T g$, where $\alpha$ is the thermal expansion coefficient and $\delta T$ is the temperature anomaly to a reference state. Thus, the scaling relationships given by Equations (\ref{equat10}) and (\ref{equat11}) are, respectively,
\begin{equation}
    {fV \over D} \sim {g\alpha {\Delta T \over L}},
\label{equat12}
\end{equation}
\begin{equation}
    {V \over L} \sim {W_E \over D},
\label{equat13}
\end{equation}
where $V$ is the horizontal velocity scale, Ekman pumping velocity $W_E$ scales the vertical velocity under wind forcing (Equation (\ref{equat9})), $\Delta T$ is the characteristic meridional temperature difference across the thermocline, and $D$ and $L$ are vertical and horizontal distance scales, respectively. These scaling relationships yield the depth of the wind-influenced thermocline:
\begin{equation}
    D \sim {({{W_E f L^2} \over {g\alpha \Delta T }})^{1/2}}.
\label{equat14}
\end{equation}

Equation (\ref{equat14}) is commonly used to scale the wind-influenced thermocline depth on Earth. But it should be noted that, Equation (\ref{equat14}) is obtained by assuming the zonal flow speed and meridional flow speed are in the same magnitude (Equations (\ref{equat12}) \& (\ref{equat13})), which is always true for Earth \citep{vallis2019essentials}. However, this assumption might not be necessarily true in our simulations due to the lack of exposed continents (e.g., Figure \ref{fig3}(c) \& (d)). In our simulations, $U$ and $V$ follow the scaling as
\begin{equation}
    f V \sim A_h {{ U} \over {L^2}},
\label{equat15}
\end{equation}
where $A_h$ is the horizontal viscosity coefficient. The U-V relationship is given by the balance between the horizontal dissipation and the Coriolis force in the interior ocean when the horizontal viscosity is relatively strong (Appendix \ref{sec: momentum_budget}),
\begin{equation}
    -f v = A_h {{{\partial}^2 u} \over {\partial y^2}}.
\label{equat16}
\end{equation}
And, Equations (\ref{equat12}), (\ref{equat13}), and (\ref{equat15}) yield
\begin{equation}
    D \sim f L^2 {({{W_E} \over {A_h g\alpha \Delta T }})^{1/2}}.
\label{equat17}
\end{equation}
Equations (\ref{equat17}) and (\ref{equat9}) suggest that the wind-influenced thermocline depth is proportional to the square root of the magnitude of wind stress curl, which is roughly consistent with our results (Figure \ref{fig5}(c)).

\begin{figure*}[ht]     
\centering
\includegraphics[scale=0.5]{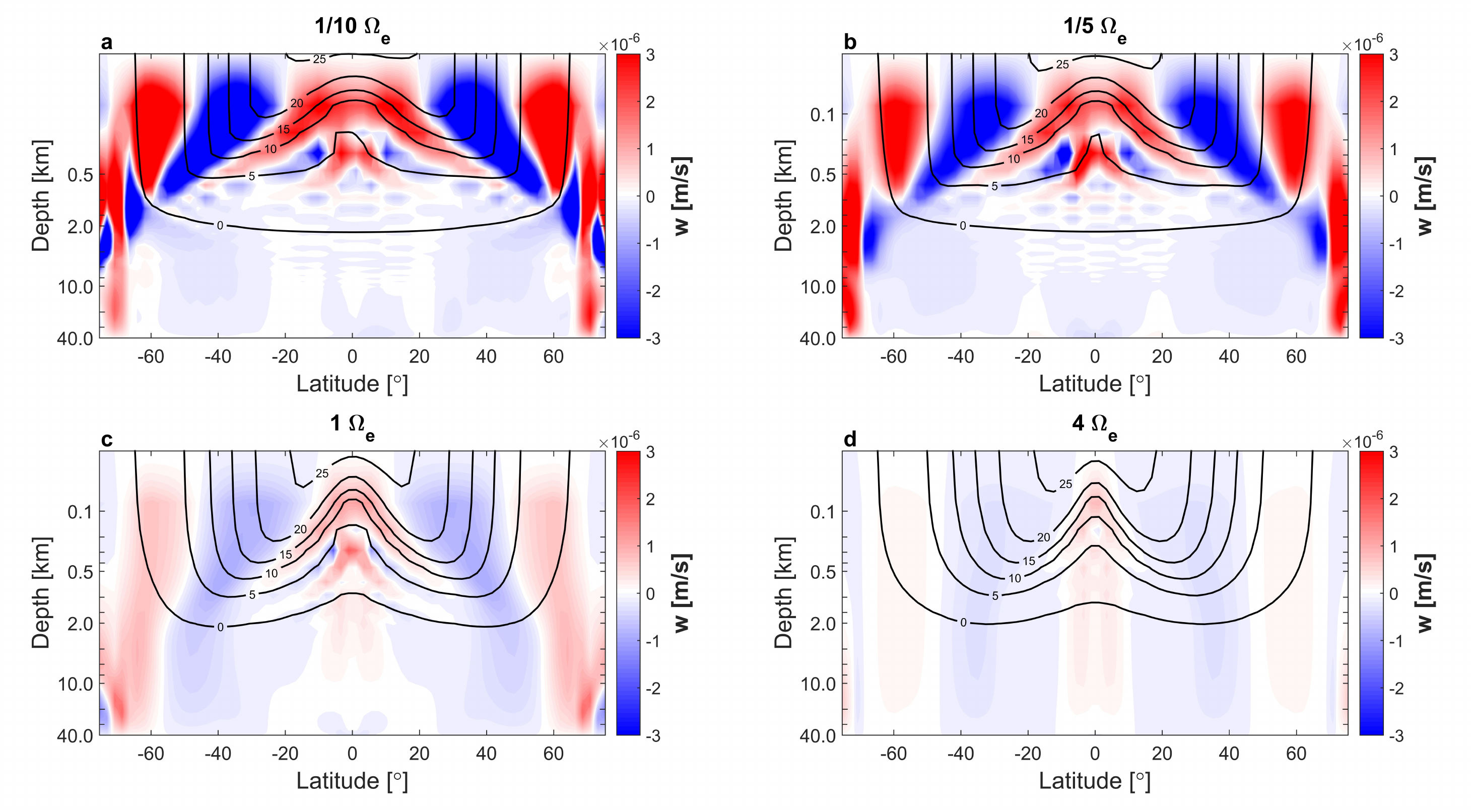}
\caption{Vertical velocity fields (color shading, m\,s$^{-1}$) and potential temperature fields (contour lines, $^{\circ}$C) under different planetary rotation rates. From (a)-(d), the planetary rotation rates are 1/10, 1/5, 1, and 4 times the Earth's rotation rate, respectively. The y-axis is nonlinear. Note that the abnormal vertical velocities in the northern and southern boundaries are possibly induced by the solid walls used there in the model. These solid walls do not exist in the real oceans of water-rich exoplanets, and the vertical velocities there are also unreliable.\label{fig6}}
\end{figure*}

\subsubsection{The effect of varying planetary rotation rates\label{subsubsec:rotation}}

Equation (\ref{equat17}) indicates that the wind-influenced thermocline depth increases with planetary rotation rate and scales as ${\Omega}^{1/2}$, which is roughly consistent with the 30$^{\circ}$S--30$^{\circ}$N mean values in our simulations (red points and red line in Figure \ref{fig5}(f)). As planetary rotation rate increases, Ekman subduction near the equator becomes weaker (Equation (\ref{equat9}) \& Figure \ref{fig6}) and less cold water is brought up, which results in smaller vertical temperature gradients and a deeper thermocline in the low-latitude regions (Figure \ref{fig5}(e)). Thus, within the main upwelling regions of 30$^{\circ}$S--30$^{\circ}$N, the mean thermocline depth increases with planetary rotation rate, especially for slowly rotating cases.

With increasing planetary rotation rate, both the upwelling motion near the equator and $\pm$60$^{\circ}$ and the downwelling motion near $\pm$30$^{\circ}$ become weaker. Thus, the thermocline depth becomes deeper in the upwelling regions but shallower in the downwelling regions under larger rotation rates (Figure \ref{fig5}(e)). At the same time, the latitudinal width of the upwelling zone at low latitudes decreases, and that of the downwelling regions extends equatorward at the same time (Figure \ref{fig5}(e) \& Figure \ref{fig6}). The narrowing of the upwelling regions and the widening of the downwelling regions under larger rotation rates suggest that the opposite change of thermocline depth may cancel out with each other and makes no net contribution to the global-mean value of the thermocline depth as shown in Figures \ref{fig5}(d) and \ref{fig5}(f)). And, the mean thermocline depth of 30$^{\circ}$S--30$^{\circ}$N also increases at a slower rate under high rotation rate cases due to the extended downwelling motion toward the equator (red points in Figure \ref{fig5}(f)).  

\subsubsection{Other parameters\label{subsubsec:other parameters}}

\begin{figure*}[ht]     
\centering
\includegraphics[scale=0.55]{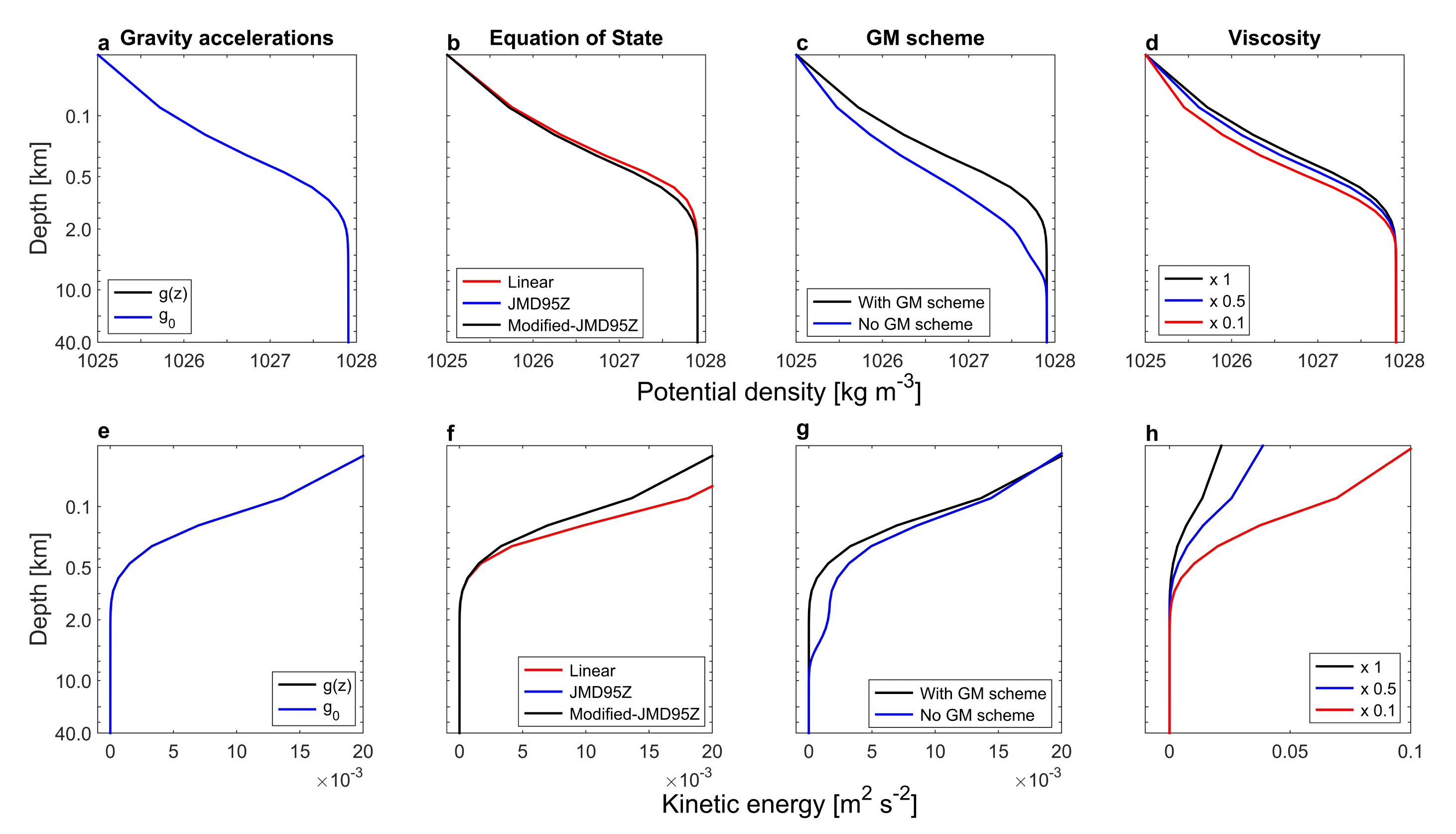}
\caption{Sensitivity of the wind-influenced thermocline depth to several parameters. Top row: global-mean potential density profiles (kg\,m$^{-3}$); bottom row: global-mean kinetic energy profiles (m$^2$\,s$^{-2}$). From left to right: the gravitational acceleration, equations of state, GM scheme, and viscosity, respectively. In (a) \& (e), and (b) \& (f), the blue line and the black line overlap with each other. Note the y-axis is nonlinear. \label{fig7}}
\end{figure*}

The sensitivity of the wind-influenced thermocline depth to other parameters is also examined (Figure \ref{fig7}). In the control case, the variation of gravitational acceleration with depth is considered and the EoS used is the modified `JMD95Z' (Table \ref{tab2}). Neglecting the depth dependence of gravitational acceleration (Figure \ref{fig7}(a) \& (e)) or varying the EoS (Figures \ref{fig7}(b) \& (f)) do not result in an obvious difference in the thermocline depth. This is probably because the wind-influenced thermocline is always confined within the upper ocean. However, turning off the GM scheme has a significant impact on the thermocline depth. As Figures \ref{fig7}(c) \& (g) show that the ocean becomes warmer and the thermocline reaches deeper when the GM scheme is turned off, extending from a depth of $\sim$2 km to $\sim$8 km. This result is consistent with previous studies, for example, \cite{danabasoglu1995sensitivity}, who suggested that the thermocline is more diffused than observed if a specified horizontal diffusion rather than the GM scheme is used. We also test the sensitivity of the wind-influenced thermocline depth to viscosity. We decrease both horizontal and vertical viscosity coefficients by a factor of 2 and 10, respectively, and find that the the kinetic energy increases and the thermocline depth becomes deeper, but the variation is limited. For instance, the thermocline deepens from a depth of $\sim$0.5 km to $\sim$2 km when the viscosity is decreased by a factor 10 (Figure \ref{fig7}(d) \& (h)).

~\

\begin{figure*}[ht]     
\centering
\includegraphics[scale=0.47]{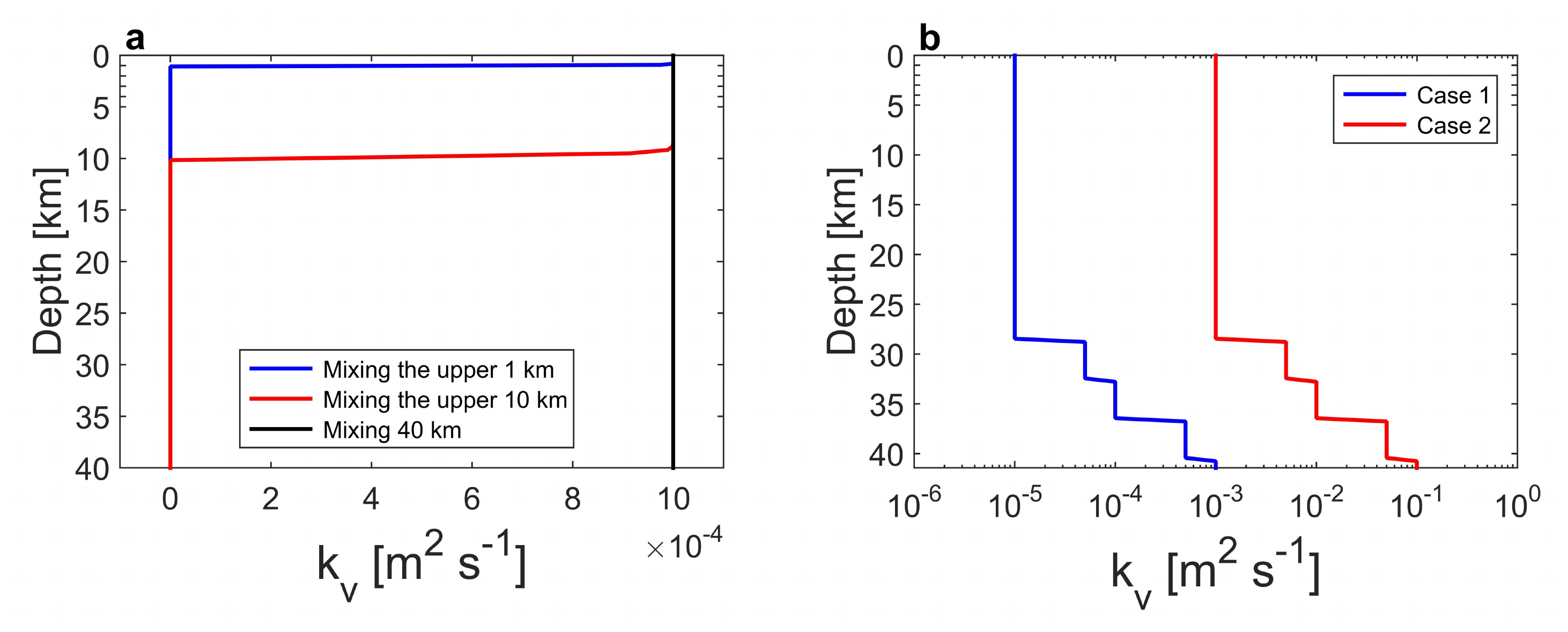}
\caption{Profiles of vertical eddy diffusivity in two series of simulations. (a) Upper-ocean enhanced mixing cases: the vertical diffusivity is 10$^{-3}$ m$^2$\,s$^{-1}$ at the surface and decreases to zero at 1 km (blue line) or at 10 km (red line); the black line represents a case in which mixing is uniform throughout the whole ocean with a diffusivity of 10$^{-3}$ m$^2$\,s$^{-1}$ for comparison. (b) Deep-ocean enhanced mixing cases: the vertical diffusivity starts to increase at a depth of $\sim$30 km and increases by two orders of magnitude at the ocean bottom; the diffusivity increases from 10$^{-5}$ to 10$^{-3}$ m$^2$\,s$^{-1}$ (blue line) or from 10$^{-3}$ to 10$^{-1}$ m$^2$\,s$^{-1}$ (red line). \label{fig8}}
\end{figure*}

\subsection{ Thermocline Depth in Mixing-dominated System \label{subsec:thermal-driven}}

Deep ocean circulation (or called thermohaline circulation) is mainly driven by turbulent mixing (e.g., \citealt{vallis2019essentials}), and the mixing-influenced thermocline depth on Earth is demonstrated to be sensitive to the vertical (or more precisely, diapycnal) eddy diffusivity \citep{bryan1987parameter,hu1996sensitivity,vallis2000large}. To examine the relationship between vertical mixing and the thermocline depth on water-rich exoplanets, we design three series of simulations: upper-ocean enhanced mixing, vertically uniform mixing, and deep-ocean enhanced mixing. In the upper-ocean enhanced mixing cases, the vertical mixing with a diffusivity of 10$^{-3}$ m$^2$\,s$^{-1}$ is allowed in the upper 1 km and 10 km, respectively (see blue and red lines in Figure \ref{fig8}(a)); for comparison, a case in which a strong, uniform vertical mixing throughout the whole ocean is also carried out (black line in Figure \ref{fig8}(a)). 
The effect of varying the strength of vertical mixing is also examined in the vertically uniform mixing scenarios, and the diffusivity ranging from 10$^{-6}$ to 2$\times$10$^{-3}$ m$^2$\,s$^{-1}$ is tested. 
In the deep-ocean enhanced mixing cases, the vertical diffusivity is enhanced in the bottom 10 km of the ocean by two orders from the background value (Figure \ref{fig8}(b)). This series of case is designed to mimic the effect of bottom topography on enhancing the mixing in the deep ocean \citep{st2002estimating,nikurashin2013overturning}. Note that in the real oceans of Earth, regions of enhanced mixing are approximately between the sea floor and the above 0.5 or 1 km \citep{BuoyancyForcingbyTurbulenceaboveRoughTopographyintheAbyssalBrazilBasin,GlobalPatternsofDiapycnalMixingfromMeasurementsoftheTurbulentDissipationRate}. While in our simulations, we extend this decay scale to 10 km to more clearly see the effect of bottom topography if it have.
The effect of varying planetary rotation rates on the thermocline depth under high-diffusivity is also examined in the vertically uniform mixing scenarios. The planetary rotation rate tested ranges from 1/30 to 8 times the Earth's rotation rate. 
Again, for simplicity, salinity is set to be uniform everywhere and only the density forcing induced by temperature difference is included in these simulations. 

\begin{figure*}[ht]     
\centering
\includegraphics[scale=0.57]{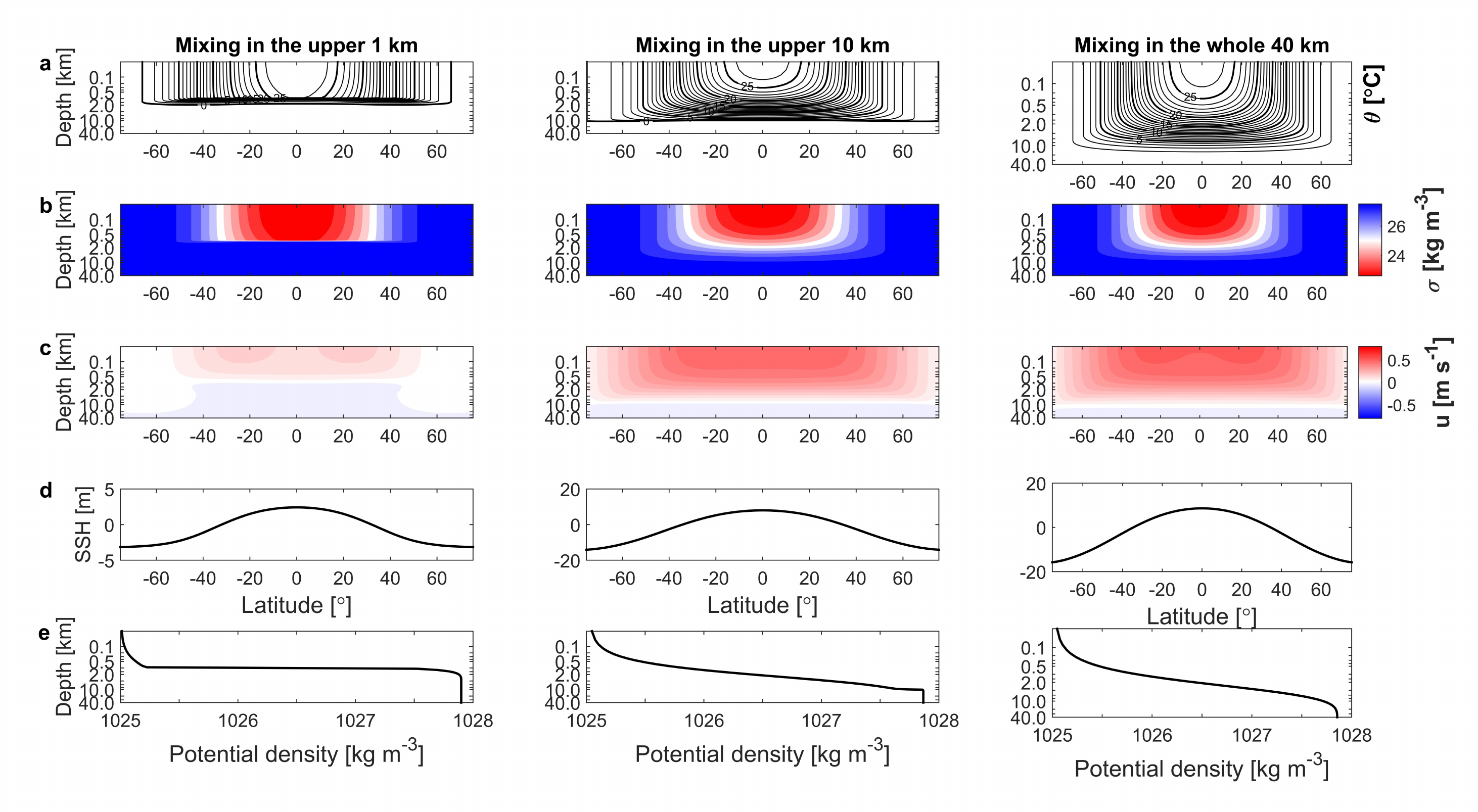}
\caption{Equilibrated states of three simulations corresponding to the three profiles of vertical diffusivity in Figure \ref{fig8}(a). Mixing with a diffusivity of 10$^{-3}$ m$^2$\,s$^{-1}$ in the upper 1 km (left column), in the upper 10 km (middle column), and in the whole 40 km (right column). (a) Potential temperature with an contour interval of 1 $^{\circ}$C; (b) potential density (kg\,m$^{-3}$, a value of 1000 kg\,m$^{-3}$ has been subtracted); (c) zonal current (m\,s$^{-1}$); (d) sea surface height (m); (e) profiles of global-mean potential density (kg\,m$^{-3}$). A horizontal diffusivity of 1000 m$^2$\,s$^{-1}$ is used and the GM scheme is turned off. Wind forcing is not included in these three experiments. Y-axis is nonlinear in (a), (b), (c), and (e). \label{fig9}}
\end{figure*}

There is no stratification where there is no mixing (Figure \ref{fig9}). Without wind forcing, the distribution of the equilibrated sea surface temperatures with latitude is similar to the imposed sea surface temperature forcing (see Figure \ref{fig2}(c)). In the ocean interior, the vertical mixing acts to diffuse heat from the upper, warmer layer downward to the lower, colder layer. Thus, the isotherms (isopynals) extend downward to the ocean interior (Figure \ref{fig9}(a) \& (b)). When the vertical mixing is restricted within the upper 1 or 10 km, the isotherms (isopycnals) are also forced to be restricted in the upper 1 or 10 km, respectively. Below those critical layers, the ocean is filled with the downwelling cold water from high latitudes and the potential density is vertically uniform (Figure \ref{fig9}(e)).  When the strong vertical mixing is allowed throughout the entire ocean, the heat is diffused downward freely (right column of Figure \ref{fig9}). Even though, the isotherms extend to only about 10 km and can not reach the bottom of the ocean (40 km) when the vertical diffusivity is as large as 10$^{-3}$ m$^2$\,s$^{-1}$, which is nearly the largest diffusivity found in small local regions of Earth's oceans \citep{GlobalPatternsofDiapycnalMixingfromMeasurementsoftheTurbulentDissipationRate}.

The spatial distribution of SSH with latitude is also similar to that of the imposed sea surface temperature forcing (Figure \ref{fig9}(d)). But, there is no SSH minimum near the equator as that was found in the wind-driven circulation cases (Figure \ref{fig3}(e)); this is due to the lack of Ekman subduction and upwelling of cold water there. 
The meridional gradient of SSH increases as the vertical mixing acts deeper in the ocean (Figure \ref{fig9}(d)). This is because the meridional gradients of sea surface temperature are also diffused deeper when the heat is diffused downward deeper from ocean surface to ocean interior.
The zonal currents at the sea surface are in geostrophic balance and are directly determined by the meridional gradients of SSH. Thus, the surface zonal currents become stronger as the vertical mixing is allowed to act deeper (Figure \ref{fig9}(c)). The zonal currents at the sea surface are eastward in the upper ocean, and there are two maxima near $\pm$20$^{\circ}$ latitudes due to the largest SSH gradients there. 

\begin{figure*}[ht]     
\centering
\includegraphics[scale=0.55]{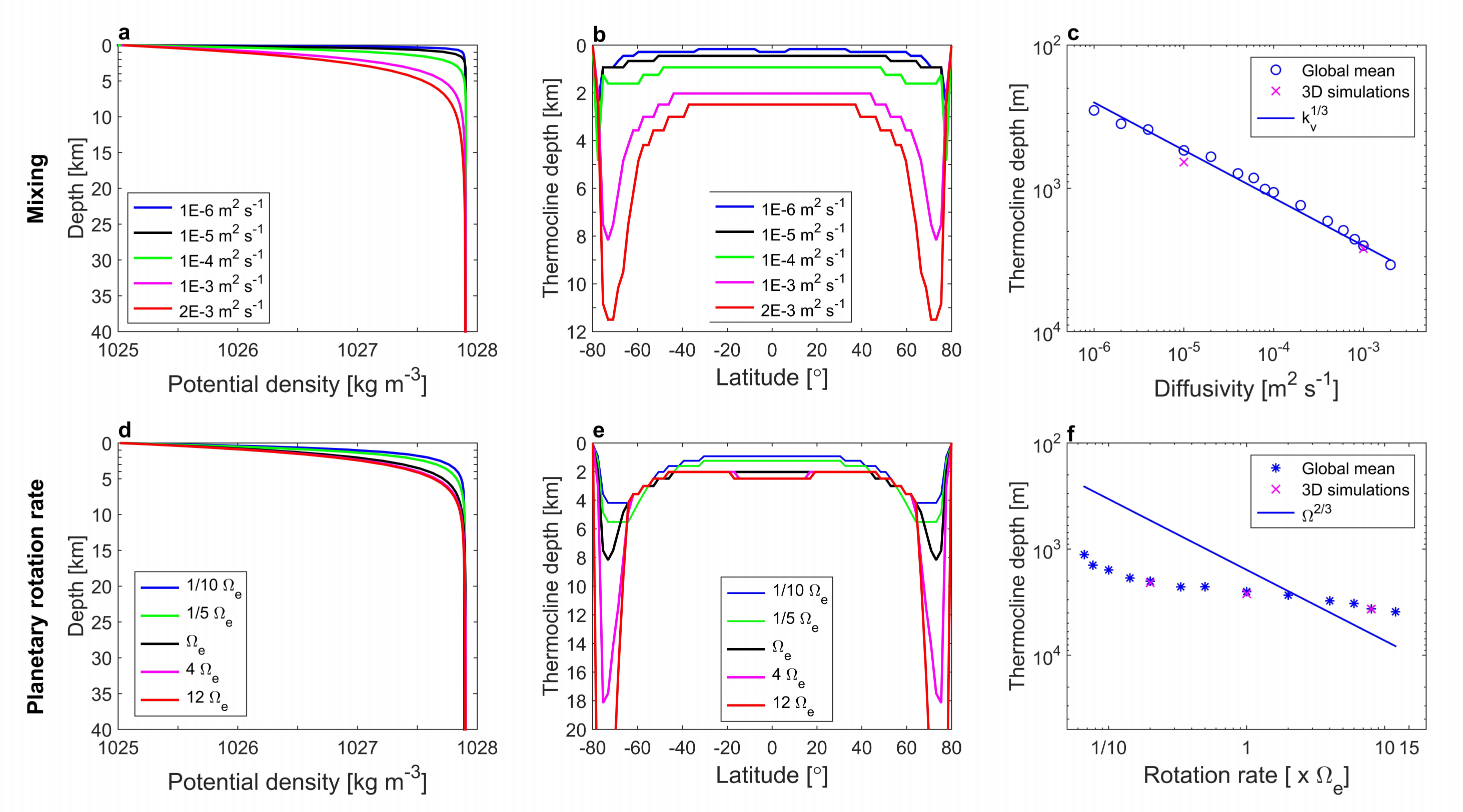}
\caption{Sensitivity of the thermocline depth to vertical eddy diffusivity (top row) and planetary rotation rate under high-diffusivity (bottom row). Upper panels: (a) global-mean potential density profiles; (b) thermocline depth as a function of latitude; (c) comparisons between numerical results and theoretical scaling relation (Equation (\ref{equat23})): blue points are global-mean thermocline depth of numerical results, blue line shows the scaling relation, and purple crosses are results of 3D simulations. Lower panels: same as the upper panels, but for different planetary rotation rates under a high, uniform diffusivity of 10$^{-3}$ m$^2$\,s$^{-1}$. Wind forcing is not included in these simulations. \label{fig10}}
\end{figure*}

The thermocline depth becomes deeper with stronger mixing (Figure \ref{fig10}(a)-(c)). When the diffusivity is low, the thermocline is shallow everywhere. As the vertical diffusivity increases, heat from the upper warm water is diffused downward deeper to the interior ocean, which weakens the vertical stratification and then deepens the thermocline at all latitudes (Figure \ref{fig10}(a) \& (b)). 
It is also worth noting that, the thermocline depth exhibits more obvious variation with latitude under high diffusivity cases (e.g., higher than 10$^{-5}$ m$^2$\,s$^{-1}$). And, the thermocline is deep in high latitudes due to the cold water sinking and weak stratification there, and it is shallower in low latitudes.
Moreover, the slope of the global-mean thermocline depth as a function of diffusivity is approximately 1/3 in the double logarithm coordinates (Figure \ref{fig10}(c)). This can be explained using simple scaling theories \citep{Welander1971b,vallis2000large}. Considering there is no wind forcing and the mixing is strong, we scale the depth of the mixind-influenced thermocline using the same thermal-wind balance (Equation (\ref{equat10})), mass continuity (Equation (\ref{equat11})), and the U-V relationship on water-rich exoplanets provided by the balance between the horizontal dissipation and the Coriolis force in the ocean interior (Equation (\ref{equat16})). 
What's more, the thermodynamic equation considering the effect of diffusive terms is needed to be included in high-diffusivity cases, given as
\begin{equation}
    {D b' \over D t}+wN^2 =k_v {{\partial}^2 b \over \partial z^2}.
\label{equat18}
\end{equation}
where $N^2={{d\overline{b}} \over {dz}}$ is the vertical gradient of the reference buoyancy and represents the vertical stratification of the system.
The thermodynamic equation obeys an advective-diffusive balance, so the scaling relationships of Eqs. (\ref{equat10}), (\ref{equat11}), {\color{red} (\ref{equat15})}, and (\ref{equat18}) become
\begin{equation}
    {fV \over \delta} \sim {g\alpha {\Delta T \over L}};
\label{equat19}
\end{equation}
\begin{equation}
    {V \over L} \sim {w \over \delta};
\label{equat20}
\end{equation}
\begin{equation}
    fV \sim A_h {{ U} \over {L^2}};
\label{equat21}
\end{equation}
\begin{equation}
    {w \Delta T \over \delta} \sim {k_v \Delta T \over {\delta}^2};
\label{equat22}
\end{equation}
where $\delta$ is the depth of the mixing-influenced thermocline, and $w$ is no longer the imposed wind-driven Ekman velocity but internally determined. Assuming the vertical temperature/buoyancy variations across the thermocline are comparable to its meridional variations, the mixing-influenced thermocline depth is
\begin{equation}
    \delta \sim {({{k_v f^2 L^4} \over {A_h g\alpha \Delta T }})^{1/3}}.
\label{equat23}
\end{equation}
This relationship suggests a scaling of ${k_v}^{1/3}$, which is quite consistent with our numerical results (Figure \ref{fig10}(c)). Even though it is obtained from high-diffusivity assumptions, this scaling seems still applicable to low-diffusivity cases (e.g., when the diffusivity is lower than 10$^{-5}$ m$^2$\,s$^{-1}$).

\begin{figure*}[ht]     
\centering
\includegraphics[scale=0.53]{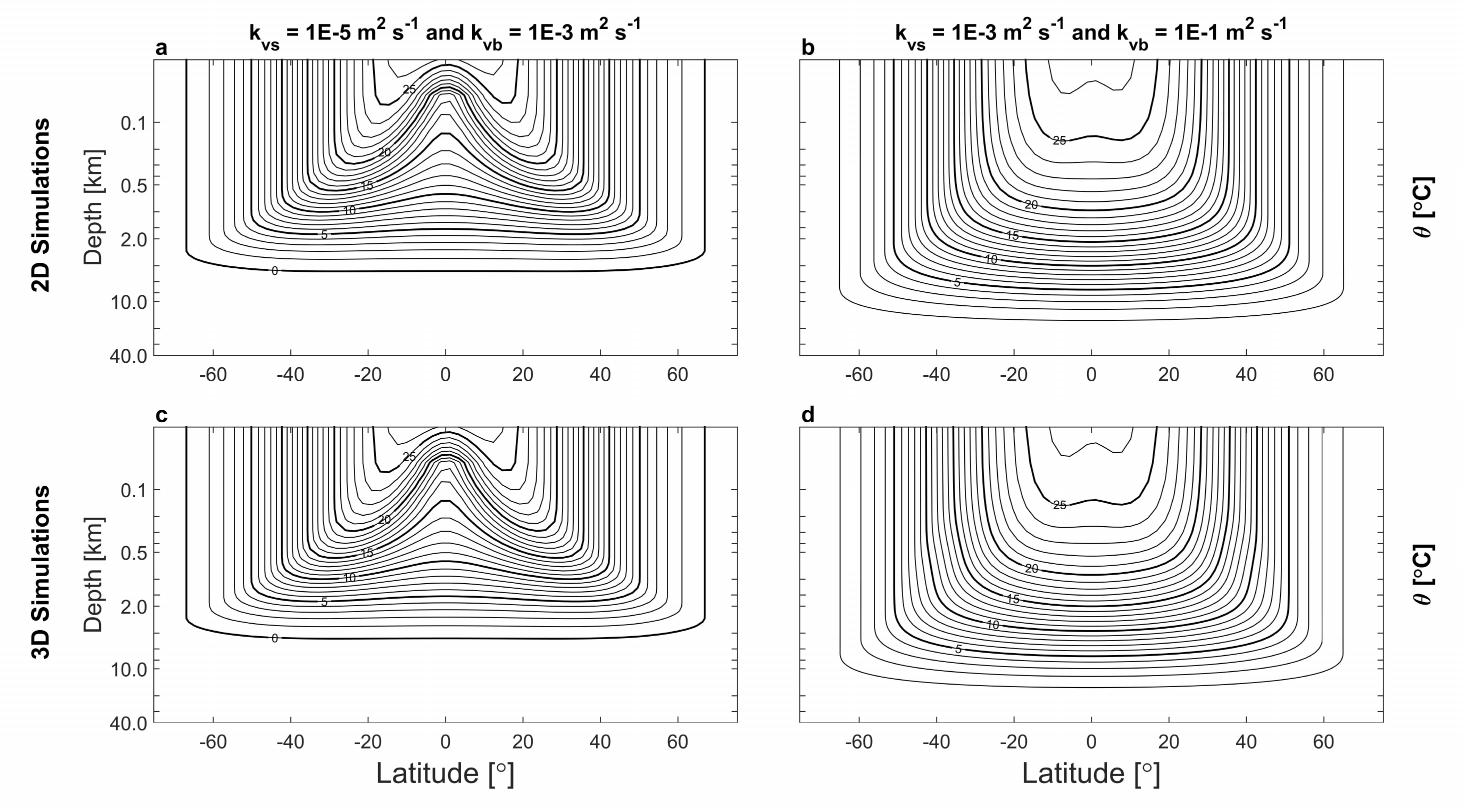}
\caption{2D simulation (top row) and 3D simulation (bottom row) results of bottom-enhanced mixing simulations. Left column: vertical diffusivity is given as the blue line in Figure \ref{fig8}(b); right column: vertical diffusivity is given as the red line in Figure \ref{fig8}(b). Wind forcing as the solid lines in Figure \ref{fig2}(a) \& (b) is included in these experiments.\label{fig11}}
\end{figure*}

The thermocline depth is also sensitive to the planetary rotation rate under high-diffusivity. As Equation (\ref{equat23}) suggests, the thermocline becomes deeper as the planetary rotation rate increases and scales as $\Omega^{2/3}$ under high-diffusivity conditions, which is roughly consistent with our global-mean results (Figure \ref{fig10}(d)-(f)). For example, the global-mean  thermocline depth increases from $\sim$1 km to $\sim$5 km when the planetary rotation rate increases from 1/10 $\Omega_e$ to 10 $\Omega_e$, where $\Omega_e$ is the Earth's rotation rate. 
The planetary rotation rate influences the mixing-influenced thermocline through affecting the internally-determined upwelling velocity. Combining Equations (\ref{equat22}) and (\ref{equat23}), the vertical velocity is
\begin{equation}
    w \sim {k_v}^{2/3} {({{A_h g \alpha \Delta T} \over {f^2 L^4}})^{1/3}},
\label{equat24}
\end{equation}
which becomes larger as the planetary rotation rate decreases. Larger upwelling velocity in the ocean interior under smaller rotation rate acts oppositely to the effect of diffusion and can result in larger vertical temperature gradients and a shallower thermocline, which is roughly consistent with the numerical results as shown in Figure \ref{fig10}(f). 

\begin{figure*}[ht]     
\centering
\includegraphics[scale=0.55]{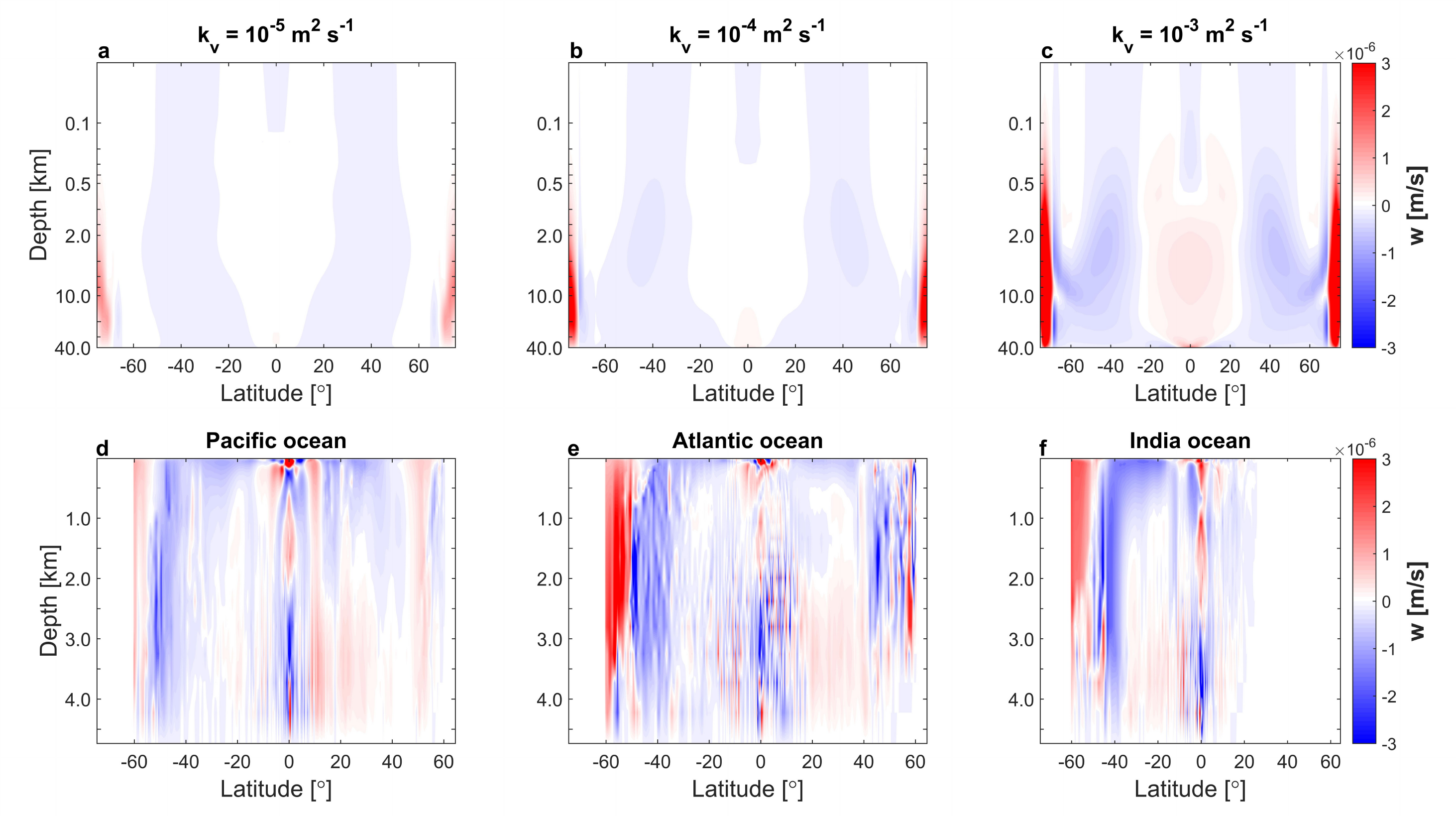}
\caption{Comparisons between the simulated mean vertical velocity (upper panels) and the mean geometric vertical velocity in the Earth's oceans (lower panels). From (a)-(c): vertical velocities (m\,s$^{-1}$) when the vertical diffusivities are 10$^{-5}$, 10$^{-4}$, and 10$^{-3}$ m$^2$\,s$^{-1}$, respectively. Note wind forcing is not included in these three simulations. The y-axis is nonlinear in (a)-(c). Similar to Figure \ref{fig6}, the abnormal vertical velocities in the northern and southern boundaries, which are possibly induced by the solid walls used there in the model, are unreliable. From (d)-(f): zonal averages of the long-term mean vertical velocities (m\,s$^{-1}$) of the Pacific, the Atlantic, and the India Oceans of the Earth, respectively. The vertical velocity data used in (d)-(f) are from the NCEP Global Ocean Data Assimilation System (GODAS, https://psl.noaa.gov/data/gridded/data.godas.html). \label{fig12}}
\end{figure*}

In the third scenario, the bottom-enhanced mixing has a limited effect on the thermocline depth (Figure \ref{fig11}). Wind stress forcing is included in this group of experiments. The effect of wind-driven Ekman subduction and the consequent cold water upwelling in the tropics is still obvious in the temperature field when the vertical diffusivity is 10$^{-5}$ m$^2$\,s$^{-1}$ in the upper 30 km (Figure \ref{fig11}(a)). When the vertical diffusivity in the upper ocean is increased to 10$^{-3}$ m$^2$\,s$^{-1}$, the strong heat diffusion dominates the temperature field and the effect of wind-induced cold water upwelling in the tropics is no longer evident (Figure \ref{fig11}(b)). In the meantime, heat is diffused downward deeper to the interior ocean as the vertical diffusivity increases, which results in a deeper thermocline. However, the thermocline is still restricted within the upper 10 km when the diffusivity in the bottom 10 km of the ocean is increased to as large as 10$^{-1}$ m$^2$\,s$^{-1}$ (Figure \ref{fig11}(b)). 
Again, results of 3D simulations are consistent with the results of the zonally symmetric 2D simulations (Figure \ref{fig11}(c) \& (d)).

The sea floor bathymetry, which is not explicitly included in our simulations, not only could affect the strength of mixing but also the viscosity at the bottom of the ocean. In our simulations, the bottom topography-induced friction is parameterized with a linear bottom drag at the sea floor. And, simple tests suggest that the influence of varying the magnitudes of the drag coefficient on the depth of the thermocline is small (figures not shown).

~\

\section{Conclusions and Discussions} \label{sec:summary}
In this paper, we investigate the thermocline depth on water-rich exoplanets where the ocean depth can reach tens of to hundreds of kilometers and there is no any exposed continent, using the global ocean model MITgcm. Our main conclusions are as follows:

1). The depths of the thermocline in our simulations are restricted within the upper ocean in most cases. In all our simulations, the maximum depth that the  thermocline can reach is about 10 km. 
Due to the lack of exposed continent, the scaling theories for the thermocline depths in both wind-dominated system and mixing-dominated system are a little different from that used on Earth. And, the scaling theories are roughly consistent with our numerical results.

2). The wind-influenced thermocline depth is mainly determined by the wind forcing and becomes deeper with larger wind stress curl. The influence of planetary rotation rate on the global-mean depth of wind-influenced thermocline is limited. Varying the equation of state and not considering the depth dependence of gravity acceleration have limited influences on the thermocline depth. The influence of turning off the GM scheme for mesoscale eddies is relatively large, which makes the thermocline reach a depth of $\sim$8 km from a depth of $\sim$2 km. The thermocline can become deeper as the horizontal and vertical viscosities decreases, but the change is small.

3). The mixing-influenced thermocline becomes deeper as the vertical eddy diffusivity or planetary rotation rate increases, but overall it can not reach the bottom of the ocean under proper parameters. 

Due to the lack of exposed continents and coastal upwelling motions on water-rich exoplanets, the depth of the ocean overturning circulation plays an important role in transporting the nutrients and other materials like carbon upward from the deep ocean to the upper ocean, which then affects planetary habitability. In our simulations, there are non-zero vertical velocities throughout the entire ocean under different magnitudes of planetary rotation rate, wind stress, and diffusivity (Figure \ref{fig6} and upper panels of Figure \ref{fig12}), which suggests that the ocean overturning circulation might reach the bottom of the ocean. The magnitudes of the mean vertical velocity in our simulations are comparable to or weaker than the magnitudes of the mean vertical velocity in the Earth's oceans (Figure \ref{fig12}). Thus, the ocean overturning circulation on water-rich exoplanets might reach the bottom of the ocean and help transport nutrients back to the thermocline and to the surface, which is beneficial to the biological activity and habitability on water-rich exoplanets. 

In this study, the effect of varying wind stresses is tested without considering the full interaction between the atmosphere and the ocean and the imposed wind stresses are steady, so ocean-atmosphere coupling model can be utilized to investigate the thermocline depth in future work.
The highest vertical diffusivity for the interior ocean we tested is 10$^{-3}$ m$^2$\,s$^{-1}$ or 100 times that on Earth, which is the estimated strength of vertical mixing on potentially habitable asynchronous rotating exoplanets around M-dwarfs \citep{si2021planetary}. However, this estimation is obtained based on simple scalings, and more rigorous calculations of the background vertical diffusivity using high-resolution ocean models are required for further investigations. 

The possible local volcanic activity at the bottom of the ocean is not included in our simulations. On Earth, the geothermal heat added at the sea floor by local volcanic activity is an important source for regional mixing in the deep ocean \citep{MixingandEnergeticsoftheOceanicThermohalineCirculation}, which could affect the overlying density fields and the stratification of the ocean and then influence the strength and depth of the meridional overturning circulation \citep{article,https://doi.org/10.1029/2005GL024956}. 
Thus, the possible local volcanic activity might also exert an influence on the thermocline depth on water-rich exoplanets.
What's more, the existence of high-pressure ice at the sea floor, which is an important characteristic of water-rich exoplanets, is also not included in our study. The high-pressure ice can reduce the bottom friction, and the decreased friction could result in a decreased strength of vertical mixing, which is unfavorable for the thermocline to reach deeper. The existence of the high-pressure ice can also directly inhibits the nutrient exchange between the ocean and the solid surface. Detailed simulations considering the effects of local volcanic activity or high-pressure ice are left for further investigations.

For simplicity, the effect of salinity is not included in this study. Both salinity forcing and temperature forcing at the surface can induce density differences and then drive ocean overturning circulation \citep{vallis2019essentials}. And, it is the strength of vertical mixing that largely determines the thermocline depths under both thermal-driven circulation and salinity-driven circulation. In this aspect, the effect of temperature forcing is basically similar to the effect of salinity forcing, and the results of simulations here that only consider the density changes induced by temperature forcing could be used to salinity-forcing experiments. 

In this study, our model has a relatively coarse horizontal resolution of 2.25\,$^{\circ}$ $\times$ 2.25\,$^{\circ}$ and can not resolve mesoscale or smaller eddies explicitly. 
A high-resolution or eddy-resolving ocean model is required to confirm the reliability of our results.
\cite{ASHKENAZY201793} used MITgcm with a high horizontal resolution of 100 m in a 10 km $\times$ 10 km domain with an ocean depth of 1.2 km to investigate the energy transfer from the surface ocean to the deep ocean. His results showed that the wind-induced currents are restricted within the top shallow layers except when the wind stress is temporally periodic with the Coriolis frequency, i.e., the wind forcing is resonating with the Coriolis force. We reproduce his two experiments but employ a 10-km-deep ocean and obtain the same result: the wind-induced kinetic energy is restricted within the surface layer under steady wind forcing but can be transferred to the deep ocean via the resonance between the wind forcing and the Coriolis force (figures not shown).
However, this resonance occurs only in very limited region(s) where the frequency of the wind stress is exactly equal to the frequency of the Coriolis force, so it can not have a significant effect on the conclusions shown in this study. 
~\

\begin{acknowledgments}
We are grateful to Yonggang Liu, Yongyun Hu, Xiaozhou Ruan, and Ru Chen for their helpful discussions. Jun Yang is supported by the National Natural Science Foundation of China (NSFC) under grants 42161144011 and 42075046. 

\end{acknowledgments}

\appendix
\setcounter{equation}{0}
\renewcommand\theequation{\Alph{section}\arabic{equation}}
\renewcommand\thefigure{\Alph{section}\arabic{figure}}
\setcounter{figure}{0}

\section{Nonlinear Equation of State} \label{sec: nonlinear_eos}
The nonlinear equation of state `JMD95Z' \citep{MinimalAdjustmentofHydrographicProfilestoAchieveStaticStability} as a function of salinity $S$, potential temperature $\theta$, and pressure $p$ is written as 
\begin{equation}
    \rho (S, \theta, p)=\frac{\rho (S, \theta, 0)}{1-p/K(S,\theta,p)}.
\label{equatA1}
\end{equation}
Note that $p$ is in bars (10$^5$ Pa) here. $\rho (S, \theta, 0)$ is a 15-term equation in terms of $S$ and $\theta$ and is written as
\begin{equation}
\begin{aligned}
  &  \rho (S, \theta, 0)=999.842594+6.793952\times 10^{-2}\theta-9.095290\times 10^{-3}{\theta}^2+1.001685\times 10^{-4}{\theta}^3\\
  &  -1.120083\times 10^{-6}{\theta}^4+6.536332\times 10^{-9}{\theta}^5+8.24493\times 10^{-1}S-4.0899\times 10^{-3}{\theta}S\\
  & +7.6438\times 10^{-5}{\theta}^2 S-8.2467\times 10^{-7}{\theta}^3S+5.3875\times 10^{-9}{\theta}^4S-5.72466\times10^{-3}S^{\frac{3}{2}}\\
  & +1.0227\times 10^{-4}{\theta}S^{\frac{3}{2}}-1.6546S\times10^{-6}{\theta}^2 S^{\frac{3}{2}}+4.8314\times10^{-4}S^2.\\
\label{equatA2}
\end{aligned}
\end{equation}
$K(S,\theta,p)$ is the secant bulk modulus and is a 26-term equation in terms of $S$, $\theta$, and $p$, which is written as
\begin{equation}
\begin{aligned}
  &  K (S, \theta, p)=1.965933\times 10^4+1.444304\times 10^2 \theta-1.706103\times 10^0{\theta}^2+9.648704\times 10^{-3}{\theta}^3\\
  &-4.190253\times 10^{-5}{\theta}^4 +5.284855\times 10^{1} S-3.101089\times 10^{-1} S \theta+6.283263\times 10^{-3}S{\theta}^2\\
  & -5.084188\times10^{-5}S{\theta}^3+3.886640\times 10^{-1} S^{\frac{3}{2}}+9.085835\times 10^{-3}S^{\frac{3}{2}}{\theta}-4.619924\times10^{-4}S^{\frac{3}{2}}{\theta}^2\\
  &  +3.186519\times 10^0p+2.212276\times 10^{-2}p{\theta}-2.984642\times 10^{-4}p{\theta}^2+1.956415\times 10^{-6}p{\theta}^3\\
  & +6.704388\times 10^{-3}pS -1.847318\times 10^{-4}pS{\theta}+2.059331\times 10^{-7}pS{\theta}^2+1.480266\times 10^{-4}pS^{\frac{3}{2}}\\
  & +2.102898\times 10^{-4}p^2-1.202016\times 10^{-5} p^2\theta+1.394680\times 10^{-7}p^2 {\theta}^2-2.040237\times 10^{-6}p^2S\\
  & +6.128773\times 10^{-8}p^2S{\theta}+6.207323\times 10^{-10}p^2S{\theta}^2.\\
\label{equatA3}
\end{aligned}
\end{equation}
In modified `JMD95Z', the coefficient in the $K(S,\theta,p)$ for the term $p$ is adjusted from 3.186519 to 2.5608 following `IAPWS-95' data \citep{WagnerandPrub1995} to present the pressure dependence of density of sea water in deep ocean more accurately.

\section{Momentum Budget} \label{sec: momentum_budget}

\begin{figure*}[ht]     
\centering
\includegraphics[scale=0.47]{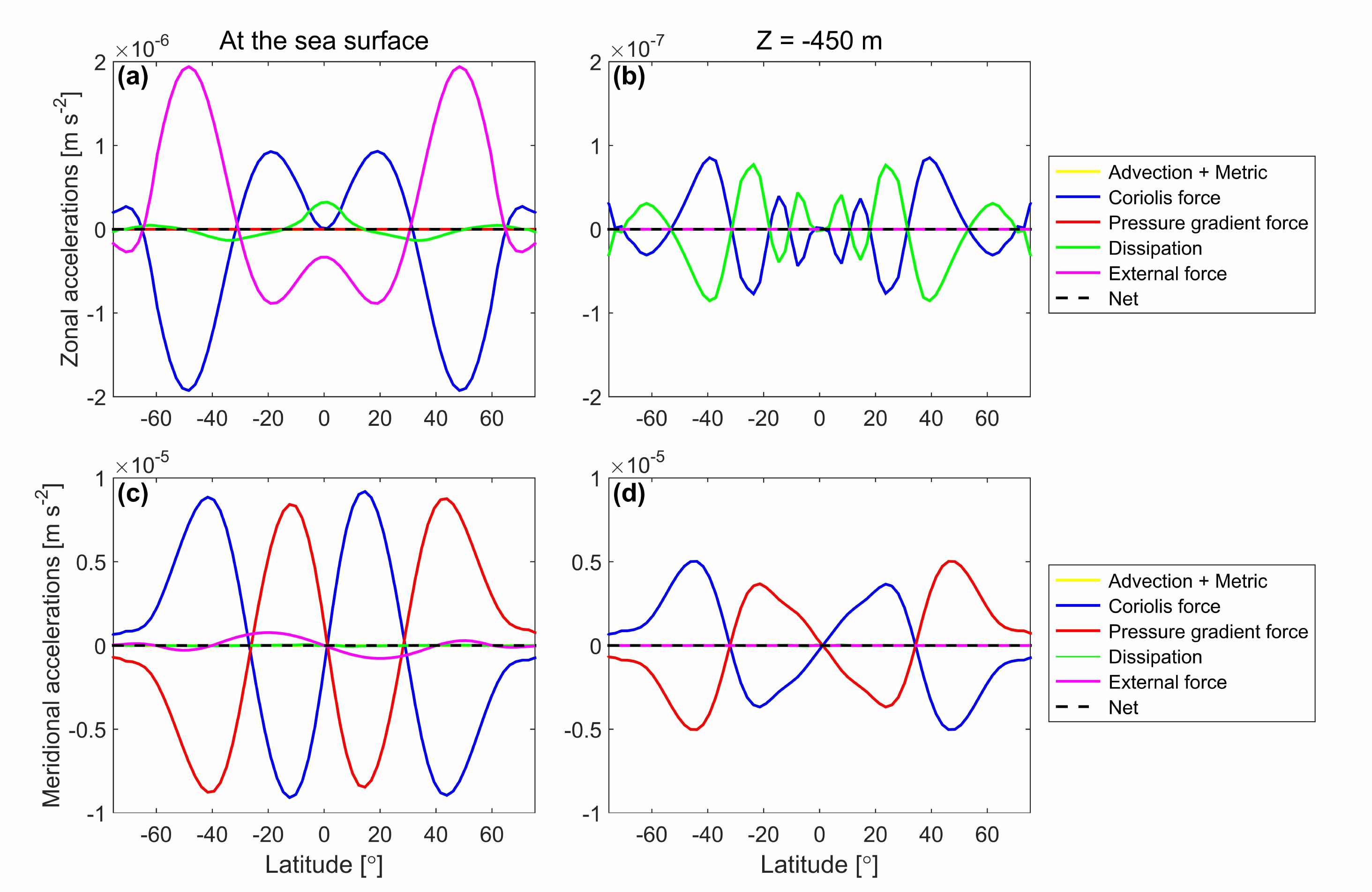}
\caption{The momentum budget at steady states of the control case. Top row: the zonal momentum budget at the sea surface (a) and at a depth of 450 m (b); bottom row: the meridional momentum budget at the sea surface (c) and at a depth of 450 m (d). The sum of advection and metric, Coriolis force, pressure gradient force, dissipation (the sum of horizontal and vertical dissipation), and external force (wind stress) terms in Eqs. (\ref{equat1}) \& (\ref{equat2}) are shown. \label{figB1}}
\end{figure*}
\begin{figure*}[ht]     
\centering
\includegraphics[scale=0.5]{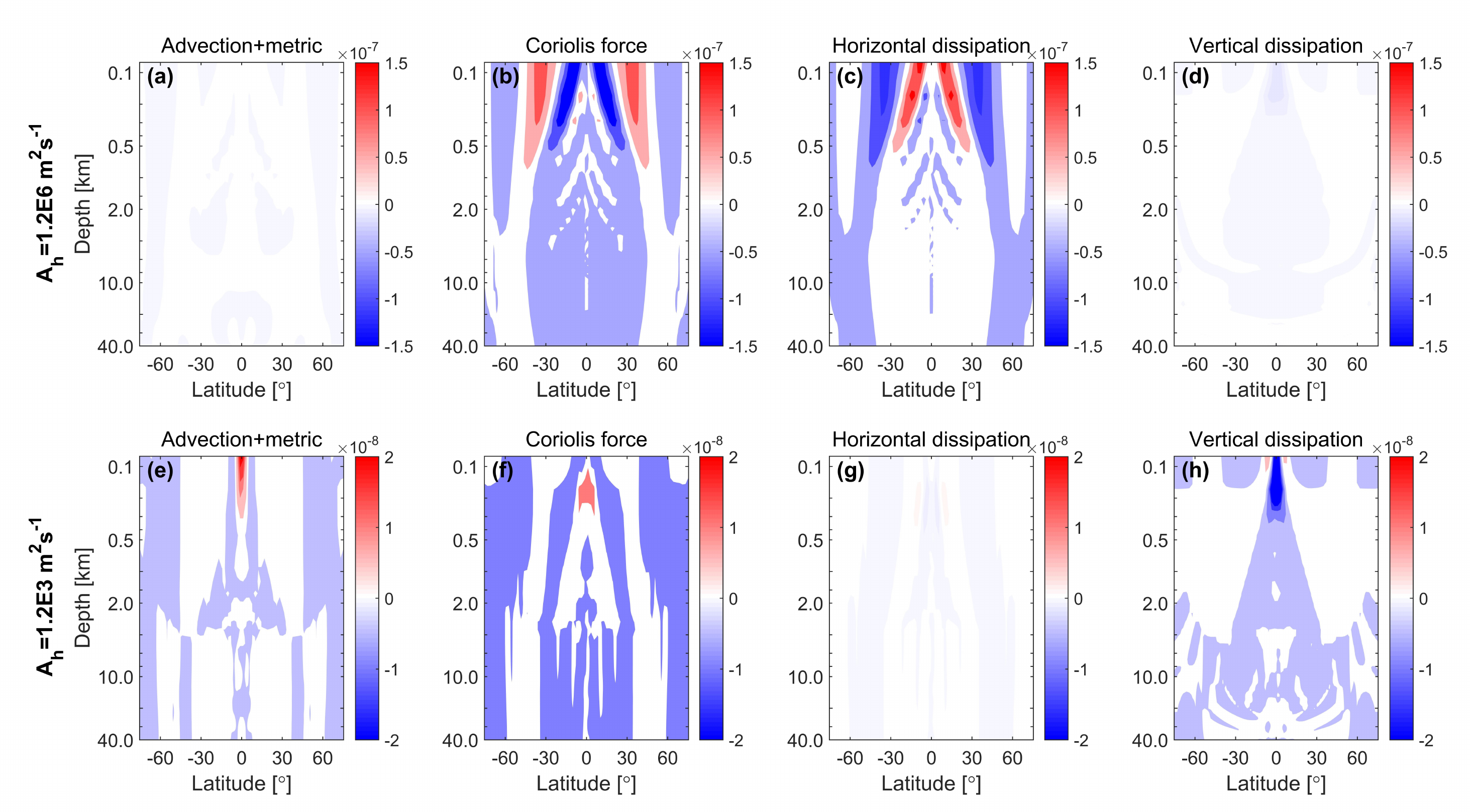}
\caption{Zonal momentum budget under different magnitudes of horizontal viscosity coefficient: A$_h$ = 1.2$\times10^6$ m$^2$\,s$^{-1}$ (upper panels) and A$_h$ = 1.2$\times10^3$ m$^2$\,s$^{-1}$ (lower panels). From left to right, it is advection$+$metric, Coriolis force, horizontal dissipation, and vertical dissipation terms of Equation (\ref{equat1}), respectively. The units are m$^2$\,s$^{-1}$. \label{figB2}}
\end{figure*}

The momentum budget at steady states of the zonnaly symmetric 2D simulations is shown here. Take the control case in Table \ref{tab2} as an example. Due to the zonal symmetry of the model, the zonal pressure gradient force is zero. In the zonal direction (Equation (\ref{equat1})), the zonal wind stress and the Coriolis force induced by the meridional currents balance with each other at the sea surface (Figure \ref{figB1}(a)); the contribution from the dissipation term at the sea surface is significant only near the equator. But, it is the dissipation term that balances with the Coriolis force in the ocean interior (Figure \ref{figB1}(b)).
Further diagnostics show that it is the horizontal dissipation that largely dominates the dissipation term and balances the Coriolis term, while the contribution from vertical dissipation is small (upper panels of Figure \ref{figB2}). The magnitude of horizontal viscosity coefficient can significantly affect the zonal momentum budget. When the horizontal viscosity is decreased by three orders of magnitude, the contributions from the other two terms, advection+metric and vertical dissipation, become dominant (lower panels of Figure \ref{figB2}).
In the meridional direction (Equation (\ref{equat2})), both the ocean surface (Figure \ref{figB1}(c)) and ocean interior (Figure \ref{figB1}(d)) are in geostrophic balance, i.e., the meridional pressure gradient force balances the Coriolis force induced by zonal currents. The contribution of the meridional wind stress is limited. The zonal-mean momentum budget of 3D simulations is quite similar to that of zonally-symmetric 2D models (figures not shown).  

It can also be seen that the accelerations in the meridional direction is one or two orders of magnitude larger than the acceleration in the zonal direction. This is because the zonal current is nearly an order of magnitude larger than the meridional current, which determines the relative magnitudes of the Coriolis force.

\setcounter{figure}{0}
\section{Baroclinic Instability in 3D simulations} \label{sec: baroclinic_insta}

\begin{figure*}[ht]     
\centering
\includegraphics[scale=0.5]{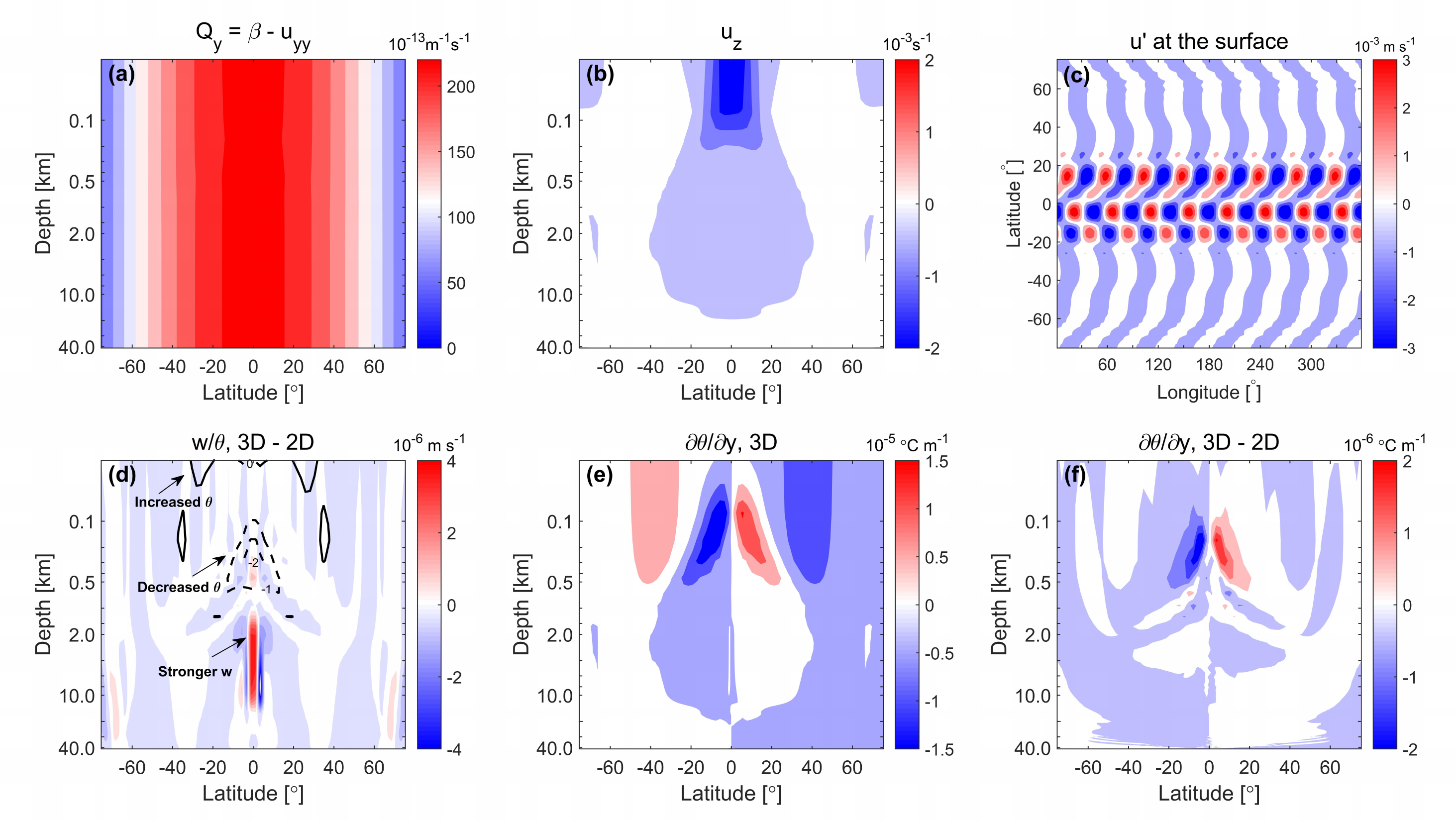}
\caption{Baroclinic instability and its effect on potential temperature in 3D simulation. Upper panels: (a) meridional gradient of potential vorticity Q$_y$ (m$^{-1}$\,s$^{-1}$); (b) vertical shear of zonal current u$_z$ (s$^{-1}$); (c) zonal current anomaly field u` (m\,s$^{-1}$) of 3D simulation. Lower panels: (d) vertical velocity differences (m\,s$^{-1}$) and potential temperature differences ($^{\circ}$C) between 3D and 2D simulations; (e) meridional temperature gradients of 3D simulation ($^{\circ}$C\,m$^{-1}$); (f) meridional temperature gradient differences between 3D and 2D simulations ($^{\circ}$C\,m$^{-1}$). \label{figC3}}
\end{figure*}

According to Charney-Stern-Pedlosky criterion, baroclinic instability might occur if the meridional gradient of potential vorticity Q$_y$ (=$\beta-\frac{{\partial}^2 u}{\partial y^2}$, where $\beta=\frac{\partial f}{\partial y}=\frac{2\Omega cos\varphi}{a}$) changes sign in the ocean interior or Q$_y$ is in the opposite sign to the vertical shear of zonal current u$_z$ at the upper boundary \citep{vallis2019essentials}. In our 3D simulations, Q$_y$ is dominated by $\beta$, which is positive everywhere and does not change sign (Figure \ref{figC3}(a)). However, u$_z$ at the upper boundary is negative and in the opposite sign to Q$_y$ within about -30$^{\circ}$--30$^{\circ}$ (Figure \ref{figC3}(b)). Thus, there should be baroclinic instability there, which can be seen from the zonal current anomaly field approximately within -30$^{\circ}$--30$^{\circ}$ both at the sea surface (Figure \ref{figC3}(c)) and in the interior ocean (figures not shown). 

With baroclinic eddy activities, compared to that of 2D simulation, the temperature in 3D simulation is slightly increased and the meridional temperature gradient is decreased near $\pm30^{\circ}$ (the edge of the area where baroclinic instability occurs) in the surface layer of the model (lower panels of Figure \ref{figC3}). But, this effect is limited possibly due to the fixed temperature forcing at the surface. There is also an evident temperature decrease in the interior ocean near the equator due to stronger upwelling in the 3D simulation, which results in larger temperature gradients there.  

\bibliography{thermocline_depth}{}

\bibliographystyle{aasjournal}

\end{document}